\documentclass{aa}

\usepackage[]{graphics}
\usepackage[]{psfig}
\usepackage[]{epsfig}
\usepackage[]{longtable}
\begin{document}

\title{Potential of the Surface Brightness Fluctuations method to measure distances 
to dwarf elliptical galaxies in nearby clusters}

\author {Steffen Mieske \inst{1,2}, Michael Hilker \inst{2} \and Leopoldo Infante \inst{1}}

\offprints {S.~Mieske}
\mail{smieske@astro.puc.cl}

\institute{
Departamento de Astronom\'\i a y Astrof\'\i sica, P.~Universidad Cat\'olica,
Casilla 104, Santiago 22, Chile
\and
Sternwarte der Universit\"at Bonn, Auf dem H\"ugel 71, 53121 Bonn, Germany
}

\date {Received 16 December 2002 / Accepted 23 January 2003}

\titlerunning{SBF distances to dEs in nearby clusters}

\authorrunning{S.~Mieske et al.}

\abstract{The potential of the 
Surface Brightness Fluctuations (SBF) method to determine the membership of
dwarf elliptical galaxies (dEs) in nearby galaxy clusters is investigated. 
Extensive simulations for SBF measurements on dEs in the $I$ band for various 
combinations of distance modulus, seeing and integration time are presented, 
based on average VLT FORS1 and FORS2 zero points.
These show that for distances up to 20 Mpc (Fornax or Virgo cluster distance), 
reliable membership determination of dEs can be obtained down to very faint 
magnitudes $-10<M_{\rm V}<-12$ mag ($\mu_0({\rm V})\simeq 25$ mag arcsec$^{-2}$)
within integration times of the order of 1 hour and with good seeing. 
Comparing the limiting magnitudes
of the method for the different simulated observing conditions we
derive some simple rules to calculate integration time and seeing needed to reach 
a determined limiting magnitude at a given distance modulus for observing
conditions different to the ones adopted in the simulations.
Our simulations show a small offset of the order of 0.15 mag towards measuring
too faint SBF. It is shown that this is due to loss of fluctuation 
signal when recovering pixel-to-pixel fluctuations from a seeing convolved image.
To check whether our simulations represent well the behaviour of real data, 
SBF measurements for a real and simulated sample of bright 
Centaurus Cluster dEs are presented. They show that our simulations 
are in good agreement with the achievable S/N of SBF measurements on real 
galaxies.}
\maketitle

\section{Introduction}
\label{introduction}
\subsection{The faint end of the galaxy luminosity function}
Dwarf elliptical galaxies (dEs) are the most numerous type of 
galaxies in the nearby universe, especially in clusters. 
This statement has been well established since the advent of CCD detectors and the 
building of telescopes with 4-8m diameter that enabled observers to detect 
low surface brightness (LSB) objects substantially fainter than the night sky. 
With the improvement of 
observing facilities, the emphasis has over the last decade switched from 
detecting faint dwarf galaxies to quantifying well their properties and 
frequencies. Most of the times dEs are investigated in galaxy clusters, 
because for a cluster the distance and therefore the approximate angular size
of candidate dwarf galaxies is known.\\
One of the most important statistical tools in investigating galaxy populations 
is the galaxy luminosity function $\Phi(M)$, describing the frequency of 
galaxies per magnitude interval. The knowledge of its logarithmic faint end slope 
$\alpha$ is very useful for testing models of galaxy formation. There are two 
fundamental steps involved in determining the faint end of $\Phi(M)$ in 
galaxy clusters: first, finding the dwarf galaxy candidates; 
second, verifying that they are cluster members.\\
For the Local Group, $\Phi(M)$ has been 
determined down to $M_{\rm V}\simeq-9$ mag 
(e.g. Mateo \cite{Mateo98}, Pritchet \cite{Pritch99}, Van den Bergh \cite{Vanden00}), 
suggesting $\alpha\simeq-1.1$. Local Group dwarf galaxies can 
nowadays readily be resolved into single 
stars with HST and/or active optics techniques. Thus, their distance can
be determined; the second step in establishing $\Phi(M)$ is quite easy to perform.
The first step, finding them, is more difficult due to their large 
angular extent and small contrast against the stars of the Milky Way. 
The latest discoveries of more and more faint dSphs 
(e.g. Armandroff et al. \cite{Armand99}, Whiting et al. \cite{Whitin99}) 
raise the question of how complete the Local Group sample is.\\
The opposite is true for nearby galaxy clusters. Here, finding candidate
dwarf spheroidals is quite straightforward when performing deep enough photometry, but
 it is impossible to resolve them into
single stars. For example, the brightest red giants of an early type galaxy at 
the Fornax cluster distance (19 Mpc, Ferrarese et al. \cite{Ferrar00}) 
have $V\simeq 29.4$ mag (Bellazzini et al. \cite{Bellaz01}). The confirmation 
of candidate dwarf spheroidals as cluster members must consequently be based on 
morphology and is therefore subject to possible confusion 
with background LSB galaxies. One depends on statistical subtraction 
of background number counts to estimate the faint end slope. Due to the generally
low number counts, the Poisson error involved in
this statistical subtraction constitutes a major source of uncertainty in determining $\alpha$,
especially in magnitude-surface brightness bins where the contribution of background galaxies
is of the order of, or higher than, that from cluster galaxies.\\
The majority of studies that have 
determined $\Phi(M)$ in nearby galaxy clusters (e.g. Sandage et al. \cite{Sandag85},
Trentham et al. \cite{Trenth01}, \cite{Trenth02a}, \cite{Trenth02b} and 
Hilker et al. \cite{Hilker03}) suggest
a logarithmic faint end slope of $-1.0<\alpha<-1.5$, being in substantial
disagreement with CDM theory (Press \& Schechter \cite{Press74}),
which predicts $\alpha\simeq-2.0$ for the initial galaxy luminosity function (Kauffman et al. 
\cite{Kauffm00}). Other authors, such as Phillips et al. 
(\cite{Philli98}) for the  Virgo cluster and Kambas et al. (\cite{Kambas00}) 
for the Fornax cluster, suggest a significantly 
steeper faint end slope of
$\alpha\simeq-2$. This large discrepancy shows that much care must be taken when 
assigning cluster membership to galaxies for which no direct distance measurement 
is available. The poisson statistics involved, especially when subtracting background 
number counts, can lead to different authors obtaining very different results for the 
same cluster.\\
Various methods for distance determination that can be applied to brighter 
galaxies outside the Local Group are not suited for the faintest dEs: 
standard candles such as SN~Ia or Cepheids are very rare; 
radial velocity measurements, if possible, are very time consuming due to the 
large fields that have to be covered.\\
\subsection{Distances to galaxies with the SBF Method}
An intriguing possibility to unambigously determine cluster membership of large 
samples of dEs in 
nearby clusters is deep wide field imaging and application of the surface brightness 
fluctuations (SBF) method. The SBF-method was first described by Tonry \& Schneider 
(\cite{Tonry88}). 
SBF are caused by the fact that on a galaxy image
the number of stars in each seeing disc is finite and therefore subject to 
statistical fluctuations.
The amplitude of these fluctuations 
relative to the underlying mean surface brightness is inversely proportional to the
number of stars. As the number of stars per unit angle increases 
quadratically with distance, the amplitude of the SBF is inversely proportional 
to distance and can therefore serve as a distance indicator.\\
As it is well known, the integrated light of an entire stellar 
population is dominated by the light emitted from the giants. In the case of an old population,
only the red giants contribute. In the case of a younger population, there is also a 
contribution from the blue super giants. In the context of the
SBF-method one can treat the whole stellar population as consisting only of stars having 
the mean luminosity weighted luminosity $\overline{L}=\frac{\int n(L)LdL}{\int n(L)dL}$
of this population. The absolute surface brightness fluctuation magnitude 
$\overline{M}$ is then defined as $\overline{M}=-2.5\times log(\overline{L})+c$ with
$c$ being the zero point of the photometric system. To obtain the distance modulus of a galaxy
with the SBF-method, one directly measures the
apparent SBF magnitude $\overline{m}$ and derives $\overline{M}$ from a distance-
independent observable, usually a colour. Measuring the apparent magnitude and deriving
its absolute value from a distance-independent observable is a procedure common to many
distance determination methods, e.g. the P-L relation for Cepheids.\\
Tonry et al. (\cite{Tonry97}, \cite{Tonry00}, \cite{Tonry01}) have carried out 
an extensive survey to measure SBF for bright elliptical and spiral galaxies in 
22 nearby galaxy groups and clusters within 40 Mpc. 
Using distances derived from cepheids or SN~Ia in the respective galaxies, 
they obtain a relation between the absolute fluctuation magnitude $\overline{M}_{\rm I}$ 
and the dereddened colour $(V-I)$, determined in the colour range $1.0<(V-I)<1.3$:\\
\begin{equation}
\overline{M}_{\rm I}=-1.74 + 4.5 \times ((V-I) - 1.15)\;{\rm mag}
\label{sbfrel}
\end{equation}
Up to now the SBF-method has only been applied 
to small samples of nearby dEs (e.g. Bothun \cite{Bothun91}, 
Jerjen et al. \cite{Jerjen98}, \cite{Jerjen00} and \cite{Jerjen01}).\\
\subsection{Aim of this paper}
In this paper, we focus on the potential of the SBF method to determine cluster 
membership of dEs in nearby clusters. We measure SBF of simulated sets of dEs 
at 3 different distance moduli between 7.5 and 43 Mpc for various combinations 
of observing times and seeing. The paper is structured as follows: 
In Sect.~\ref{degeneracy}, the impact of the age-metallicity degeneracy on SBF magnitudes
is discussed. In Sect.~\ref{simexpl}, the technical details of simulating and 
measuring SBF are described and the properties of the different sets of 
simulated dEs are shown. In Sect.~\ref{simres}, the results of the SBF
measurements for all simulated sets are presented and the limiting absolute
magnitudes for cluster membership determination is discussed. A validity check of the
simulations is presented by comparing real and simulated SBF data.
We finish this paper 
with the conclusions in Sect.~\ref{conclusions}.\\
\section{Deriving $\overline{M}_{\rm I}$ from $(V-I)$}
\label{degeneracy}
In the context of the SBF-method, the distance modulus of a galaxy is given by
the difference between apparent and absolute fluctuation magnitude
$(\overline{m}_{\rm I}-\overline{M}_{\rm I})$. To estimate the reliability of the method, 
one must know {\it both} the accuracy in measuring $\overline{m}_{\rm I}$ at the cluster 
distance and the uncertainty in deriving $\overline{M}_{\rm I}$ for a dE 
with a given $(V-I)$.\\
\begin{figure}[h!]
\vspace{-0.0cm}
\epsfig{figure=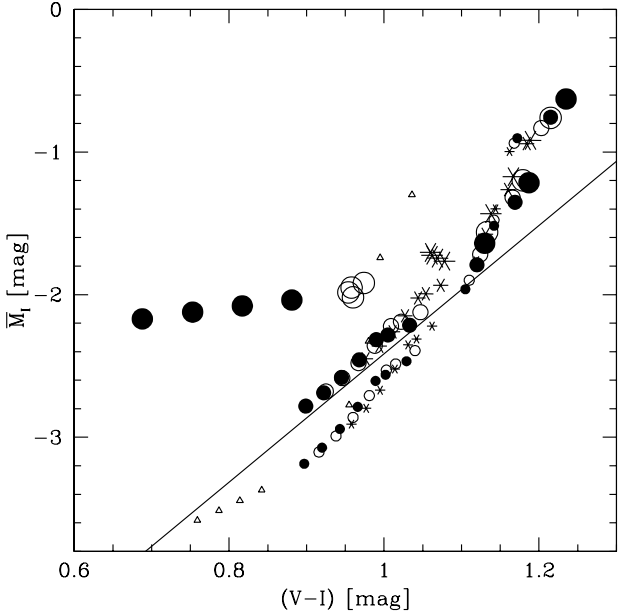,width=8.6cm,height=8.6cm}
\caption{\label{twobranch}
Theoretical absolute fluctuation magnitude $\overline{M}_{\rm I}$
  plotted vs. the colour $(V-I)$ for a set of different stellar populations, 
as taken from Worthey (1994). The larger the symbols, the older the stellar population.
Largest symbols: 17 Gyr; intermediate size symbols: 12 Gyr; small symbols: 8 Gyr 
(except for the open triangles, see below). 
{\it Solid circles}: Single 
burst populations with Fe/H between -1.7 (-1.3 for 17 Gyr) and 0 dex (left to right). 
{\it Open circles and asterisks:} 10\% and 30\% mixing of the 
single burst populations from above with a 5 Gyr population of Fe/H=0. 
{\it Open triangles}: Single burst population of 1.5 Gyr and Fe/H $\ge$ -1.3. 
The solid line is defined by Relation (\ref{sbfrel}).}
\end{figure}
In Fig.~\ref{twobranch}, theoretical values for $\overline{M}_{\rm I}$ are plotted
vs. $(V-I)$ for a set of old to intermediate age stellar populations with a wide 
range of metallicities, taken from Worthey\footnote{http://astro.wsu.edu/worthey/dial/
dial\_a\_pad.html} (\cite{Worthe94}). Equation~(\ref{sbfrel}) 
is indicated as well. Note that Tonry et al. determined equation~(\ref{sbfrel}) only for 
$1.0<(V-I)<1.3$ as they used brighter and therefore redder galaxies. The theoretical
and observed values for $\overline{M}_{\rm I}$ agree quite well, although for $(V-I)>1.15$
the slope of equation~(\ref{sbfrel}) is slightly shallower than for the theoretical values.\\
Fig.~\ref{twobranch} shows that for $(V-I)\ge1.0$, $\overline{M}_{\rm I}$ depends 
mainly on $(V-I)$ and very little on the age-metallicity combination of the 
underlying stellar population. However, for $(V-I)\le1.0$ the spread in 
$\overline{M}_{\rm I}$ at one given $(V-I)$, i.e. the effect of the age-metallicity 
degeneracy, rises significantly. At $(V-I)\simeq0.85$ this leads to an uncertainty 
of up to 0.5 mag in relating $\overline{M}_{\rm I}$ to $(V-I)$. Jerjen et al. (1998, 2000) 
have found an analogous spread of the order of 0.5 mag 
from $R$-band SBF measurements for blue dEs in the Sculptor and Centaurus A group.
Additional colour
information can help to reduce
this uncertainty; e.g., at constant 
$(V-I)=0.9$, 
$d\overline{M}_{\rm I}/d(V-K)\simeq 10$ and $d\overline{M}_{\rm I}/d(U-V)\simeq 5$
(Worthey \cite{Worthe94}). 
Precise photometry in $U$, $V$ and $I$ would therefore allow a precision of 
0.1 to 0.2 mag in deriving $\overline{M}_{\rm I}$ from theoretical models.\\
While the age-metallicity degeneracy is an important source of uncertainty for
distance measurements to blue {\it field} galaxies, it becomes useful as a 
relative age-metallicity indicator for blue {\it cluster} galaxies, as most of 
the time the cluster is separated from background/foreground galaxies by 
significantly more than 0.5 mag in distance modulus. 
Spectroscopic surveys of the Fornax cluster (Drinkwater et al. \cite{Drinkw00}, 
Hilker et al. \cite{Hilker99}) revealed a significant gap in radial velocity between
Fornax members and background galaxies corresponding to $\simeq$ 3 mag in
$(m-M)$. Confusion of a blue and young background galaxy with a blue and old 
cluster galaxy is therefore very unlikely to happen.
\\
Note that 
dIrr candidate members, to which the SBF-method is difficult to apply due to their 
irregular shape, can be distinguished morphologically in a straightforward manner
from blue young background galaxies like anemic spirals. For clusters with a 
significant fraction of dIrrs, this morphological cluster membership assignment 
can complement the assignment based on SBF-distances for the smoothly 
shaped dE candidates and allow derivation of $\Phi(M)$ for the entire dwarf galaxy
population.\\
\section{Simulating and measuring Surface Brightness Fluctuations in dEs}
\label{simexpl}
We have simulated sets of dEs in the $I$-band 
with 3 distance moduli 29.4, 31.4 and 33.4 mag, corresponding to 7.6, 19 and 48 Mpc.
This range was chosen to include distances to the more nearby groups like Leo I
(10 Mpc) as well as to the more distant clusters like Centaurus and Hydra (33 to 33.5 mag).
Note that 31.4 mag is the approximate distance modulus to Fornax and Virgo.
The integration time was 3600 seconds, the gain was 1 and the zero point 27.0 mag.
The latter value is a mean of the VLT FORS1 and FORS2 zeropoint for imaging in the 
$I$-band when including an averaged colour term and extinction coefficient.
The pixel scale was 0.2$''$/pixel, the image size 2048$\times$2048 pixel. 
For each distance modulus, a set with 0.5$''$ and 1.0$''$ seeing was simulated. 
Additionally, for 31.4 mag distance modulus and 0.5$''$ seeing, four different integration
times were adopted, namely 900, 1800, 3600 and 7200 seconds.
Note that varying the integration time from $t_1$ to $t_2$ can also be considered 
as keeping the integration time fixed and adding $2.5\log{\frac{t_2}{t_1}}$ to the zero point.\\
The photometric properties of our simulated dEs are derived from Hilker et al.'s
(\cite{Hilker03}) values found for Fornax cluster dEs. 
For a given absolute magnitude $M_{\rm V}$, the colour-magnitude- and 
surface brightness-magnitude relation from Hilker et al. (\cite{Hilker03}) is used 
to obtain $(V-I)$ and $\mu_0({\rm V})$. For example, a dE with $M_{\rm V}=-10$ mag will 
have $(V-I)=0.86$ mag
and $\mu_0({\rm V})=25.9$ mag arcsec$^{-2}$ while a dE with $M_{\rm V}=-15$ mag 
will have $(V-I)=1.04$ mag
and $\mu_0({\rm V})=22.7$ mag arcsec$^{-2}$.
An exponential intensity profile of the form 
$I(r)=I_0\times exp(-r/r_0)$ was adopted, with $r_0$ calculated from $M_{\rm V}$,
$\mu_0({\rm V})$ and the adopted distance modulus. The ellipticity was chosen as zero.
Globular Cluster (GC) systems are included, with a specific frequency of $S_N=5$
(Miller et al. \cite{Miller98}) for all galaxies, an absolute turnover magnitude 
$M_{\rm I}=-8.5$ mag (Kundu et al. \cite{Kundu01}) and the projected spatial GC density 
following the galaxy light distribution.\\
The final output of a SBF measurement procedure is the apparent surface brightness
fluctuation magnitude $\overline{m}$ (in our case $\overline{m}_{\rm I}$). 
This is equivalent to the luminosity weighted average apparent luminosity of the observed stellar
population. As $\overline{m}_{\rm I}=\overline{M}_{\rm I}+(m-M)$, with $(m-M)$ being the
adopted distance modulus of the simulated dE, 
$\overline{M}_{\rm I}$ determines the SBF amplitude at a given distance modulus.
$\overline{M}_{\rm I}$ was adopted as a function of $(V-I)$ according to Tonry's 
equation~(\ref{sbfrel}) 
for $(V-I)>$ 1.0. For $(V-I)<$ 1.0, it was decided to split the sample into two 
halves, since
it is not known whether Tonry's equation~(\ref{sbfrel}) also holds for galaxies bluewards
of this limit. For 50\% of our galaxies, $\overline{M}_{\rm I}$ was calculated according to
equation~(\ref{sbfrel}). For the other 50\% of our galaxies, $\overline{M}_{\rm I}$ was 
kept constant
at $\overline{M}_{\rm I}((V-I)=1.0))=-2.4$ mag. Doing so we acknowledge that there is a
significant age-metallicity degeneracy for $(V-I)<$ 1.0 (see Worthey's models). For the
bluest galaxies simulated, with $(V-I)\simeq0.8$ ($M_{\rm V}\simeq-8.5$ mag), this implies
a range in $\overline{M}_{\rm I}$ of about 0.9 mag between the two simulated
samples. The effect for the simulations is that, on average, the SBF signal is weaker than
if all dEs with $(V-I)<$ 1.0 were simulated according to equation~(\ref{sbfrel}).\\
To allow for varying seeing and integration times, the background field 
had to be created artificially. In a real background field obtained with VLT FORS1 in the 
$I$-band at 3000 sec integration, we fitted a power law distribution of the form 
$n(m) = A\times10^{\gamma(m_{\rm I}-m_0)}$ to the magnitude distribution of the objects 
detected by SExtractor down to the completeness limit of 
$I\simeq$ 25 mag. The fitted values were $A=14900/$degree$^2$, $\gamma=0.305$, 
$m_0=22$. According to this distribution, an artificial
object field was created with the IRAF task {\rm mkobjects} in the ARTDATA package,
with $I$=27 mag as the faint limiting magnitude. Seeing and integration time 
were chosen as needed for the simulations. The sky brightness was adopted as 19.9 
mag arcsec$^{-2}$ in $I$, which holds within 3 days before and after new moon. 
To simulate large-scale flat-field effects, the background fields were multiplied 
by the normalized sky-map obtained from applying SExtractor to a VLT FORS1 flat 
field image.\\
Into each 2048$\times$2048 pixel field, 16 dEs were implemented. For each 
set with constant seeing, distance modulus and integration time, six fields were 
created, each with a different (random) spatial distribution of the background objects.
This means that 96 dEs were simulated for each set.\\

\subsection{Simulation of Surface Brightness Fluctuations}
\label{simulation}
For each modelled pixel with a given distance $r$ to the galaxy center, 
first the number of stars corresponding to the surface brightness $\mu(r)$ 
of the exponential profile was calculated:
\begin{equation}
N_{\rm stars/pixel}(r)=10^{-0.4\times(\overline{M}_{\rm I}+(m-M)-\mu(r))}\times p^2
\label{nperpix}
\end{equation}
with $p$ being the pixel scale. Then, a random number $N^*$ was chosen within a Poisson 
distribution centered on $N_{\rm stars/pixel}(r)$. The intensity adopted at that pixel was
then defined as
\begin{equation}
I(pixel,r)=10^{-0.4\times(\mu(r)-ZP)}\times p^2 \times \frac{N^*}{N_{\rm stars/pixel}(r)}
\label{nsbf}
\end{equation}
with $ZP$ being the zero point, 27 mag in our case. 
The implementation of the SBF is apparently achieved by multiplying by 
$\frac{N^*}{N_{\rm stars/pixel}(r)}$. This means that along an isophote with radius $r$,
the intensity $I(r)$ has a pixel-to-pixel rms of 
$\frac{I(r)}{\sqrt{N_{\rm stars/pixel}(r)}}$. The image with the implemented SBF was 
then convolved with a Moffat seeing profile, which was modelled out to 7 times the FWHM. 
Finally, Poisson noise with rms=$\sqrt{I}$ was implemented. Once modelled, the 
galaxies were added onto the artificial background fields.\\
\subsection{Measurement of SBF}
\label{meassbf}
To measure the SBF of a simulated dE, the following steps were undertaken:\\
1.~Create object map with SExtractor of the whole image (containing 16 dEs)\\
2.~Mask the dEs on the object map, subtract this image from original image\\
3.~Create and subtract SExtractor sky map\\
4.~Determine and subtract local sky level by a curve of growth analysis with the IRAF-task
ELLIPSE in the ISOPHOTE package\\
5.~Model mean galaxy light with ELLIPSE using a sigma clipping algorithm to disregard
contaminating sources hidden below the galaxy light, subtract the model\\
6.~Divide resulting image by square root of the model, cut out portion where SBF 
are measured\\
7.~Mask out contaminating sources like foreground stars, background galaxies and globular
clusters.\\
8.~Calculate the power spectrum (PS) of the cleaned image\\
9.~Obtain the azimuthal average of the PS\\
10.~Fit function of the form 
\begin{equation}
\label{azimut}
P(k)=PSF(k)\times P_{\rm 0}+P_{\rm 1}
\end{equation}
to the azimuthally averaged PS. \\
$PSF(k)$ is the
PS of the seeing profile, normalized to unity at k=0. $PSF(k)$ is determined 
from a simulated star with no close neighbours by fitting a Moffat profile 
to its PS. $P_{\rm 1}$ is the white noise component, proportional to the ratio between sky
and galaxy brightness in the range where SBF were measured. It is independent of
seeing.
$P_{\rm 0}$ is the amplitude of the pixel-to-pixel surface brightness fluctuations, being the zero 
wavenumber limit of the seeing convolved pixel-to-pixel star count fluctuations, and
therefore seeing-independent, too.\\
It holds that
\begin{equation}
\label{mbarI}
\overline{m}_{\rm I}=-2.5*log(P_{\rm 0}/t_{\rm exposure}) + ZP
\end{equation}
Values at small $k$ (long wavelength) are rejected for the fit, as they are often 
considerably influenced by large-scale residuals from imperfect galaxy subtraction 
and the finite width of the image portion used to measure SBF.\\
The (seeing independent) S/N of the measurement
was defined as S/N=$P_{\rm 0}$/$P_{\rm 1}$, following Tonry\&Schneider (\cite{Tonry88}). In the
following it will be referred to as canonical S/N.
Note however that the detectability of SBF decreases with increasing seeing: by
convolving with the 
seeing the star count pixel-to-pixel fluctuations are smoothed out. Additionally,
the larger the 
seeing is, the fewer the independent data points per unit angle that are available. 
To take this dependence of the SBF
detectability on the seeing into account, we defined a modified signal to noise $S/N^*$ 
in the following way:
\begin{equation}
S/N^*=P_{\rm 0}/P_{\rm 1}\times \sqrt{N_{\rm sd}} / sf
\end{equation}
with $N_{\rm sd}$ being the number of seeing discs -- i.e. independent data points --
contained in the image portion where
SBF are measured, and $sf$ being the smoothing factor by which the seeing convolution
reduces the pixel-to-pixel fluctuations. $sf$ was 5 for 0.5$''$ seeing and 10 for 1.0$''$ 
seeing, as the effective area of one seeing disc was 25 and 100 pixels, respectively.
\subsubsection{Examples of simulated dEs}
\begin{figure*}[t!]
\epsfig{figure=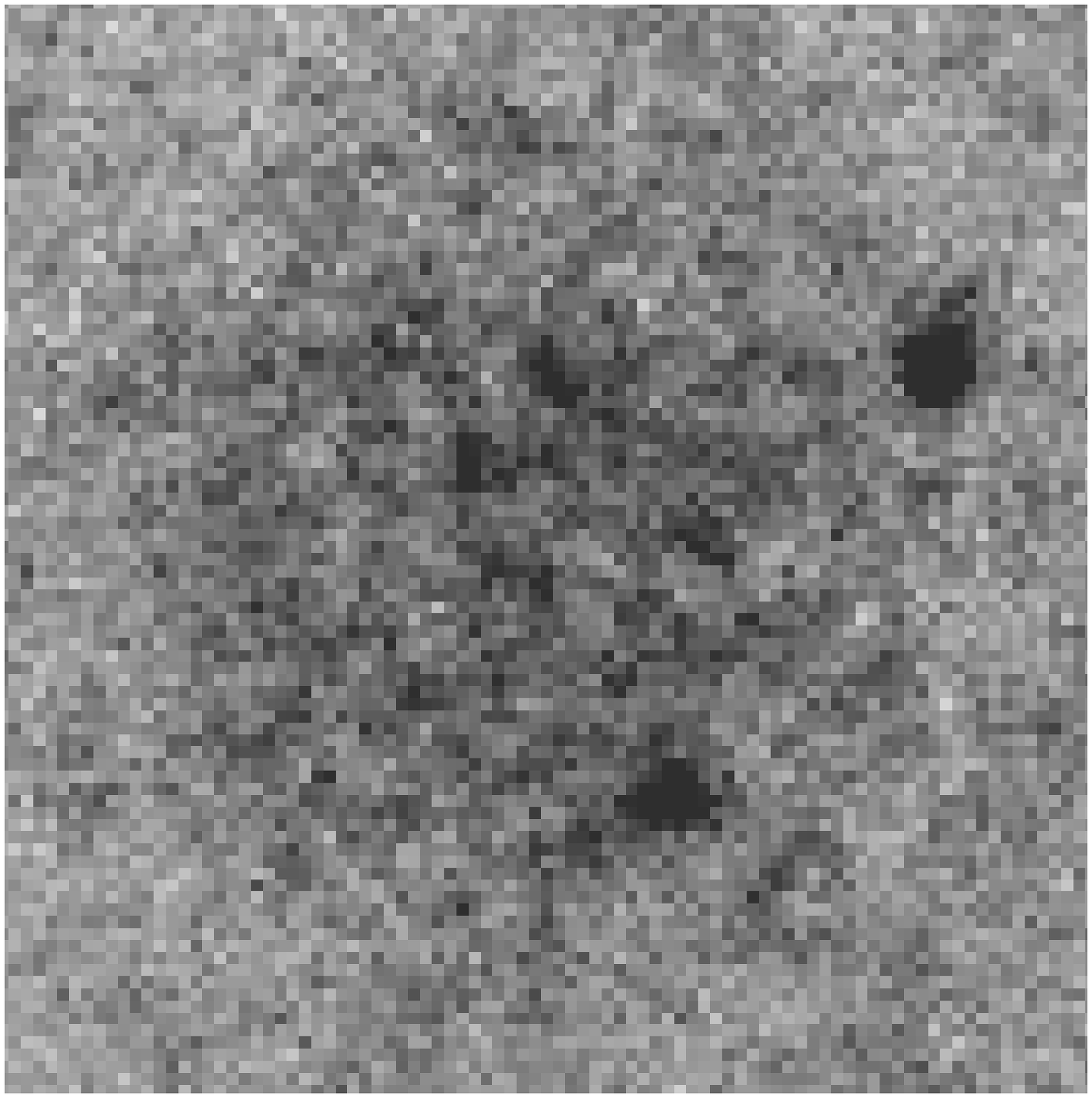,width=4.3cm,height=4.15cm}\hspace{0.4cm}
\epsfig{figure=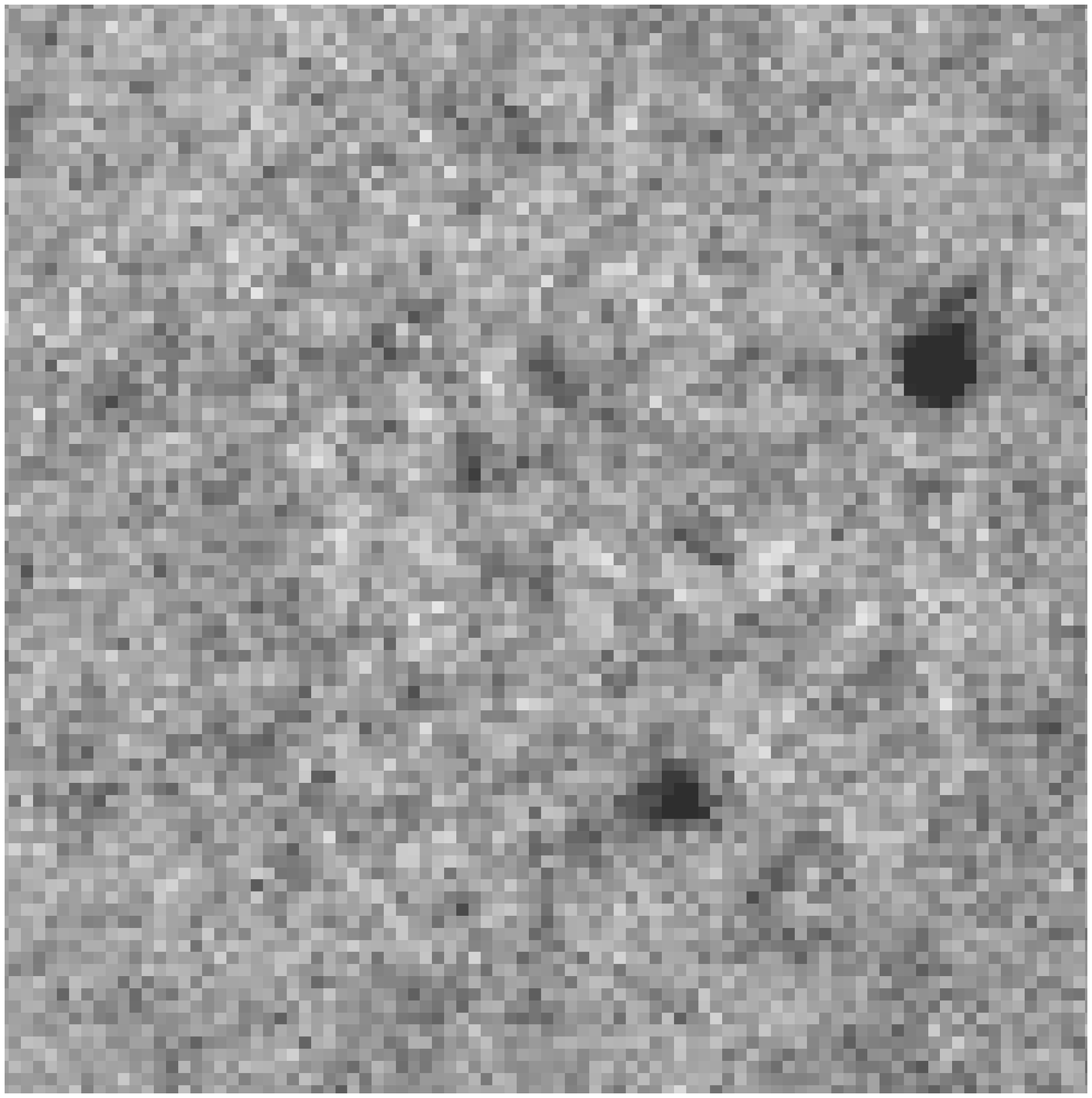,width=4.3cm,height=4.15cm}
\epsfig{figure=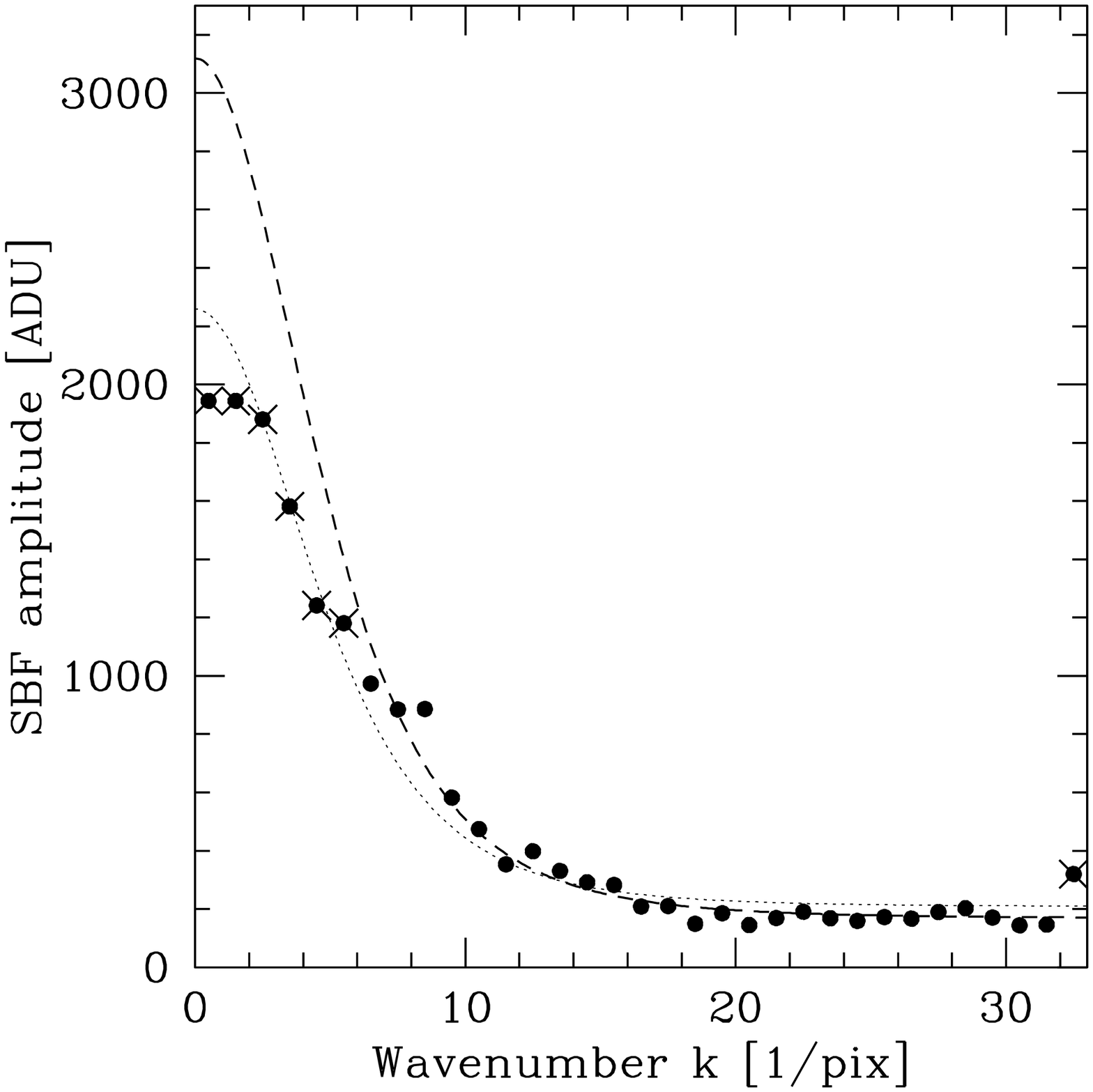,width=4.8cm,height=4.1cm}\vspace{0.05cm}\\
\hspace{0.6cm}
\epsfig{figure=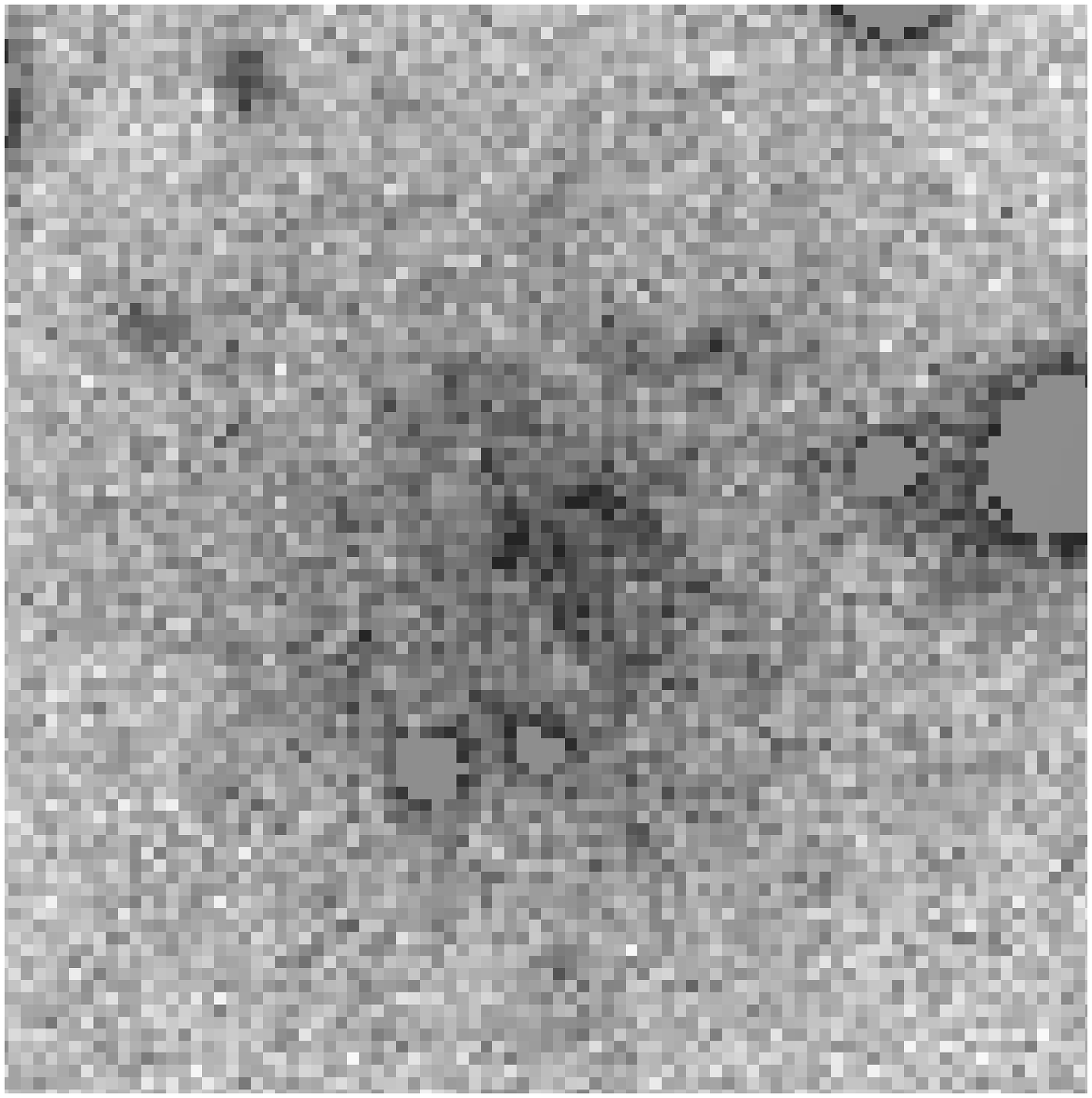,width=4.3cm,height=4.15cm}\hspace{0.4cm}
\epsfig{figure=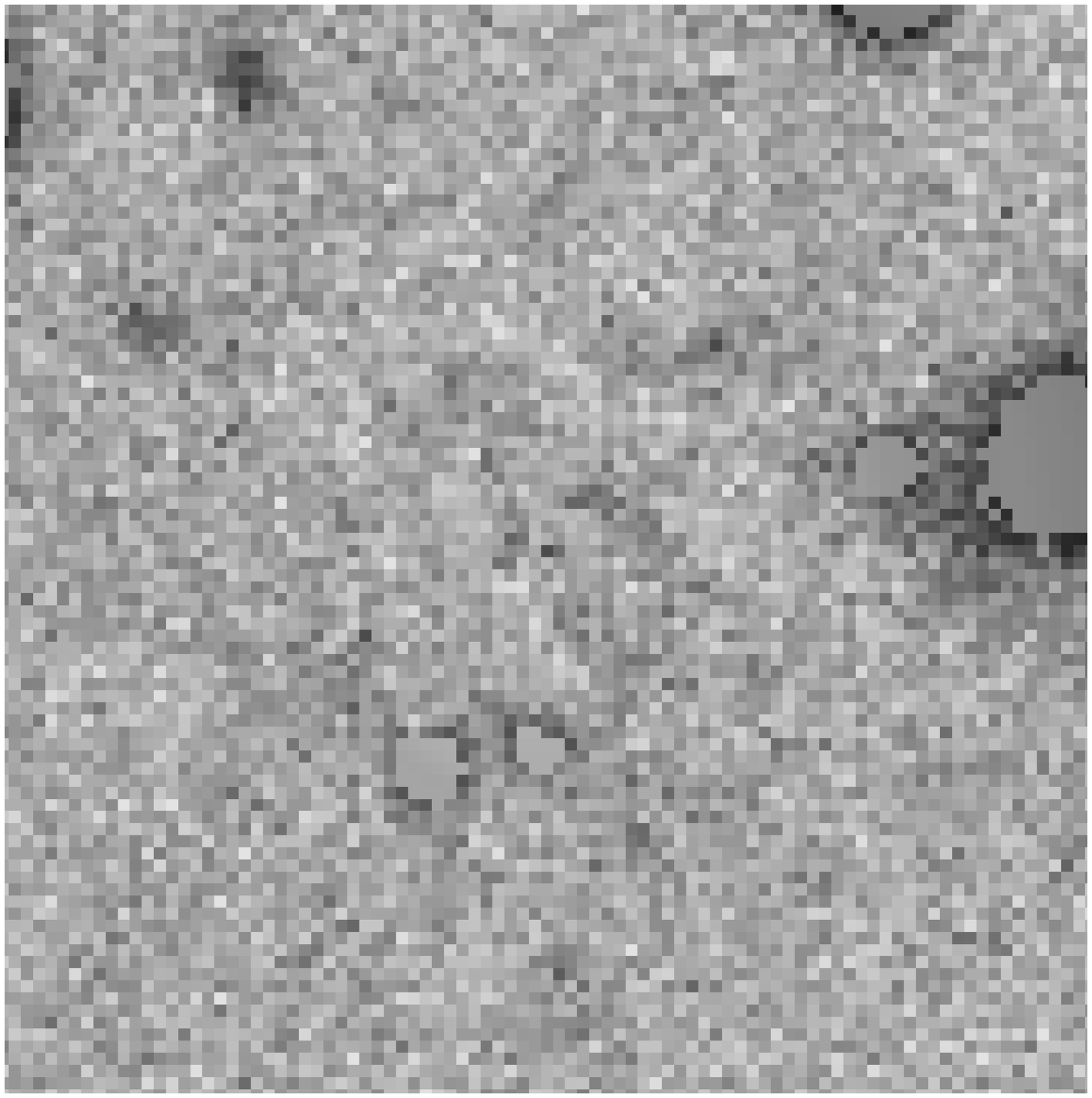,width=4.3cm,height=4.15cm}
\epsfig{figure=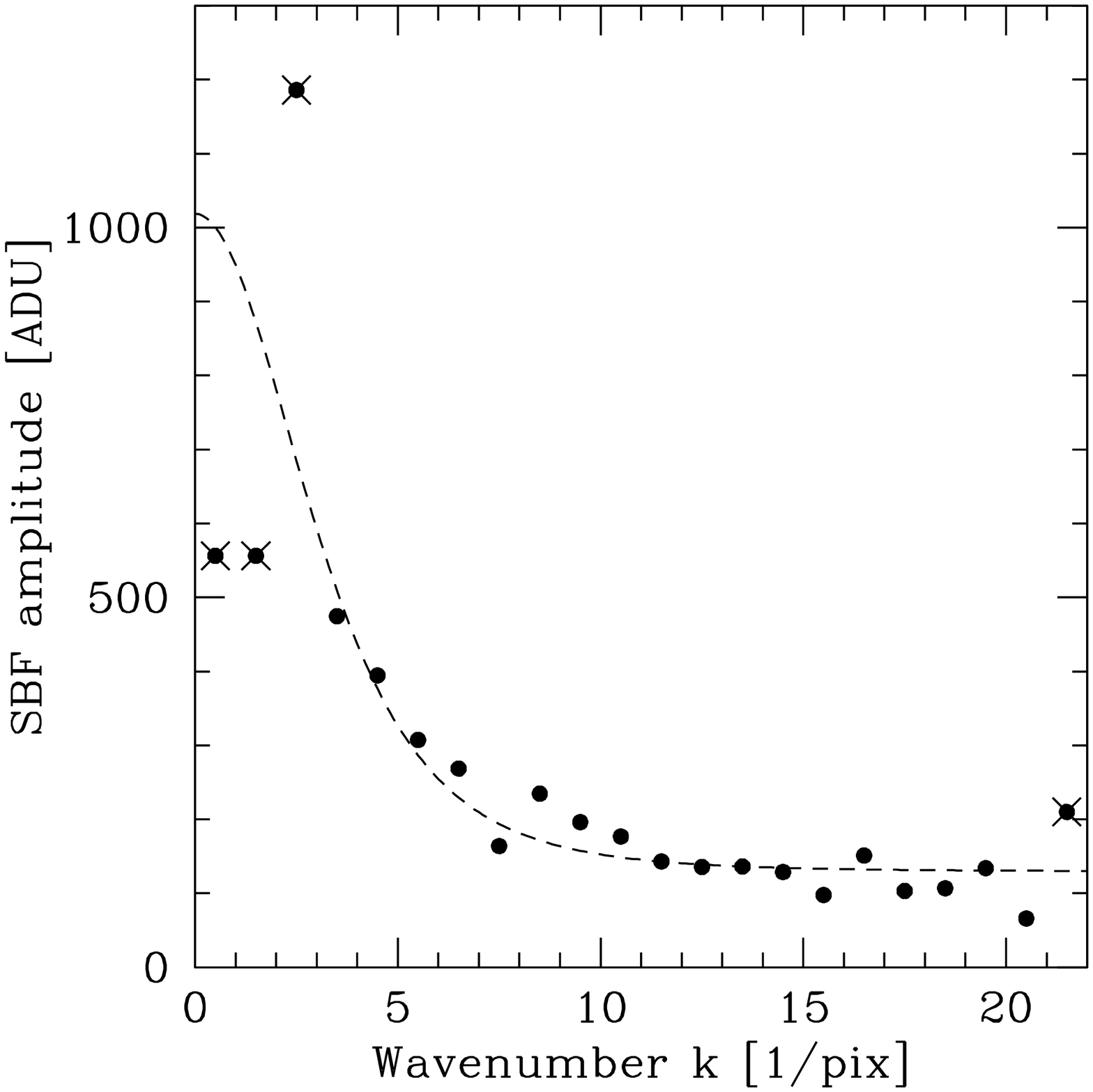,width=4.8cm,height=4.1cm}\vspace{0.05cm}\\
\hspace{0.6cm}
\epsfig{figure=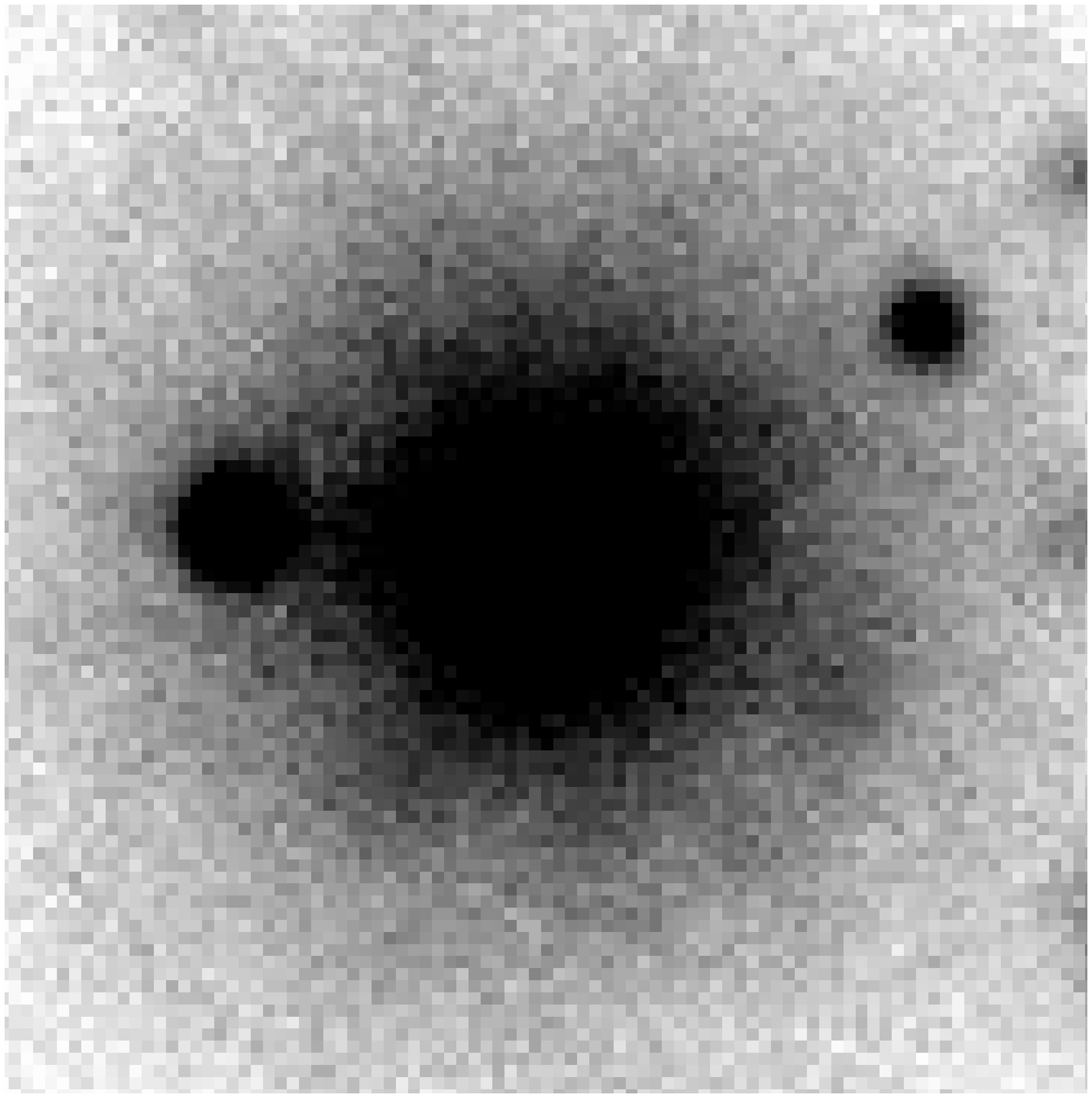,width=4.3cm,height=4.15cm}\hspace{0.4cm}
\epsfig{figure=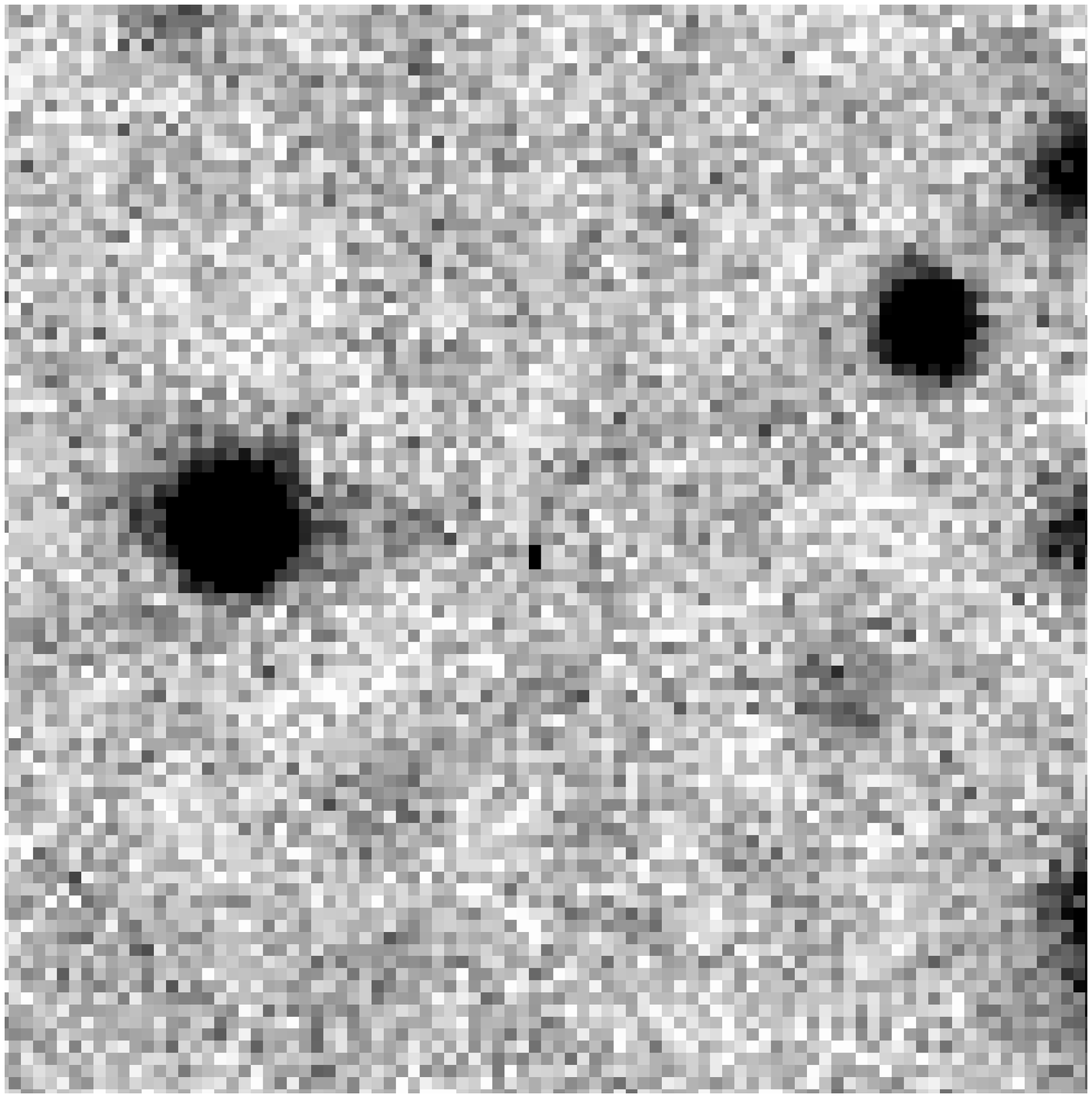,width=4.3cm,height=4.15cm}
\epsfig{figure=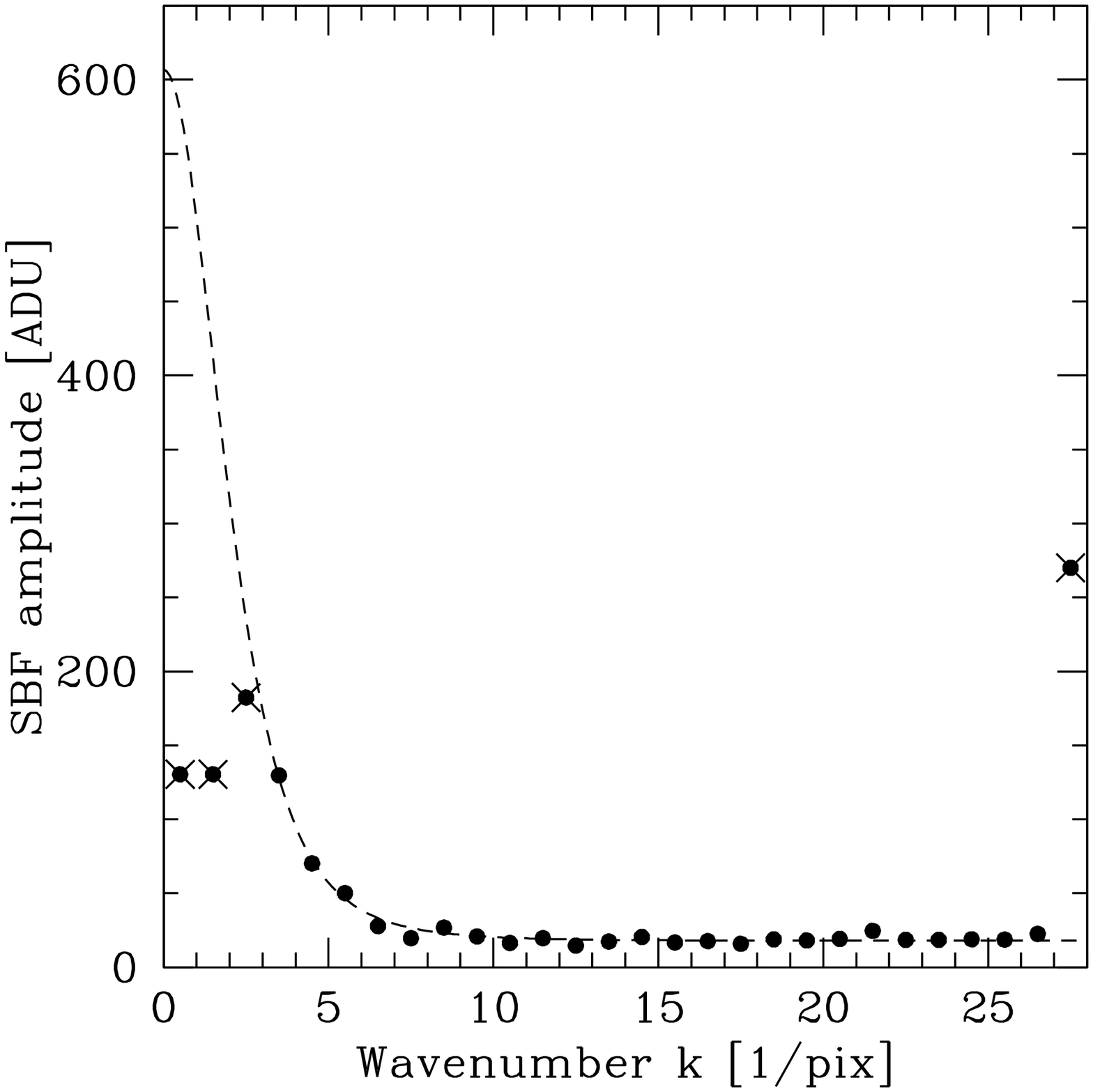,width=4.8cm,height=4.1cm}\vspace{0.05cm}\\
\hspace{0.6cm}
\epsfig{figure=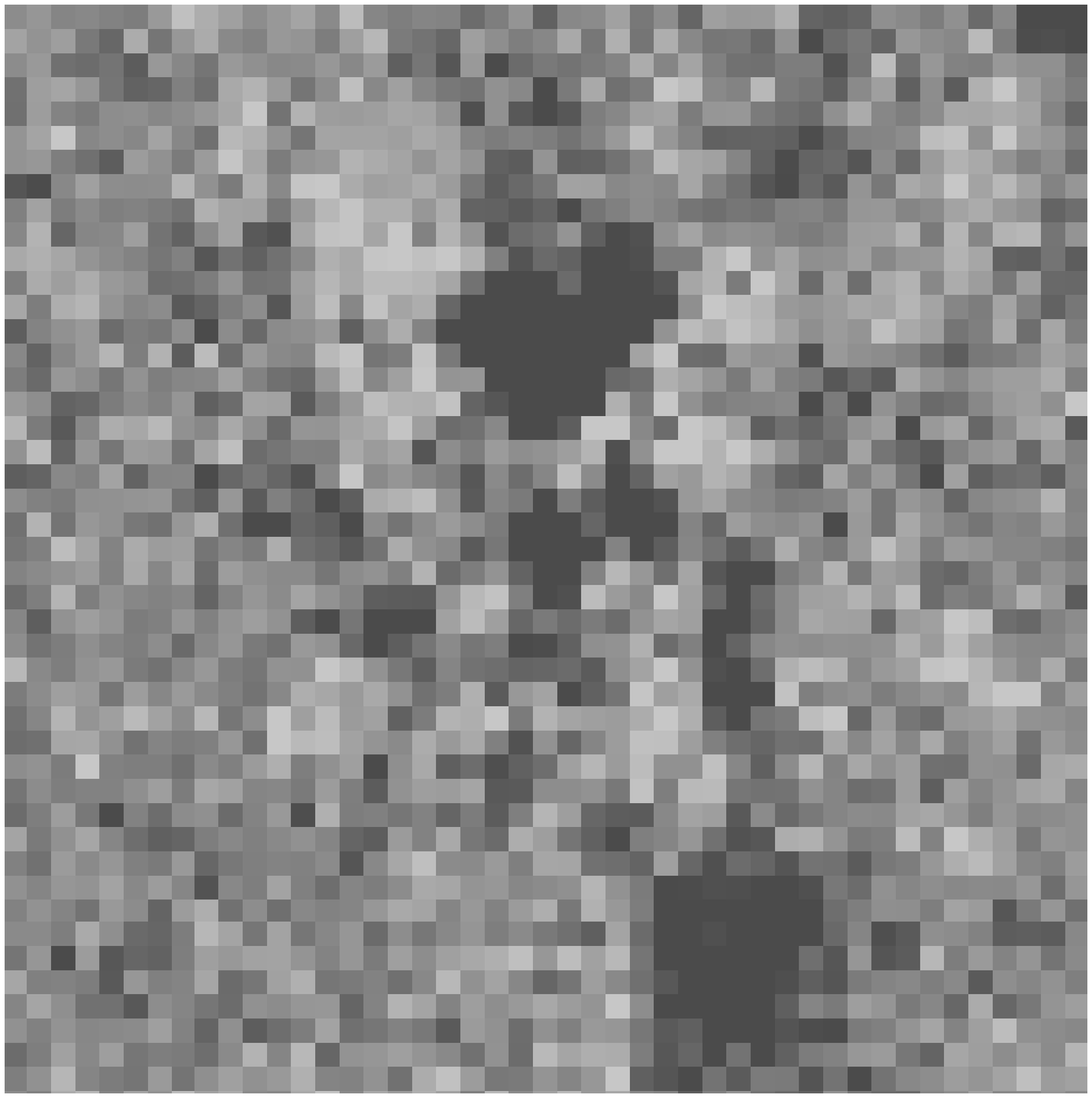,width=4.3cm,height=4.15cm}\hspace{0.4cm}
\epsfig{figure=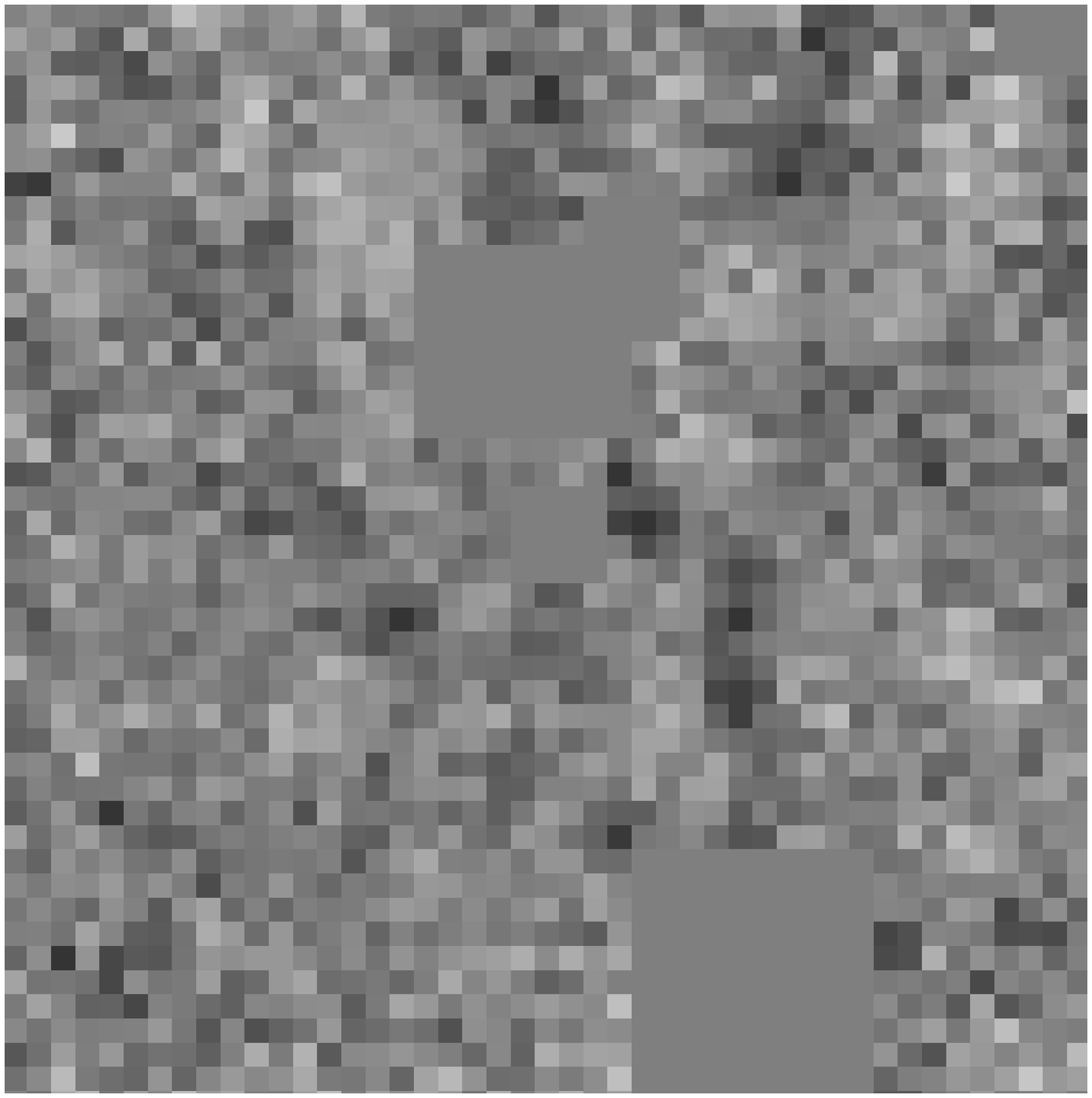,width=4.3cm,height=4.15cm}
\epsfig{figure=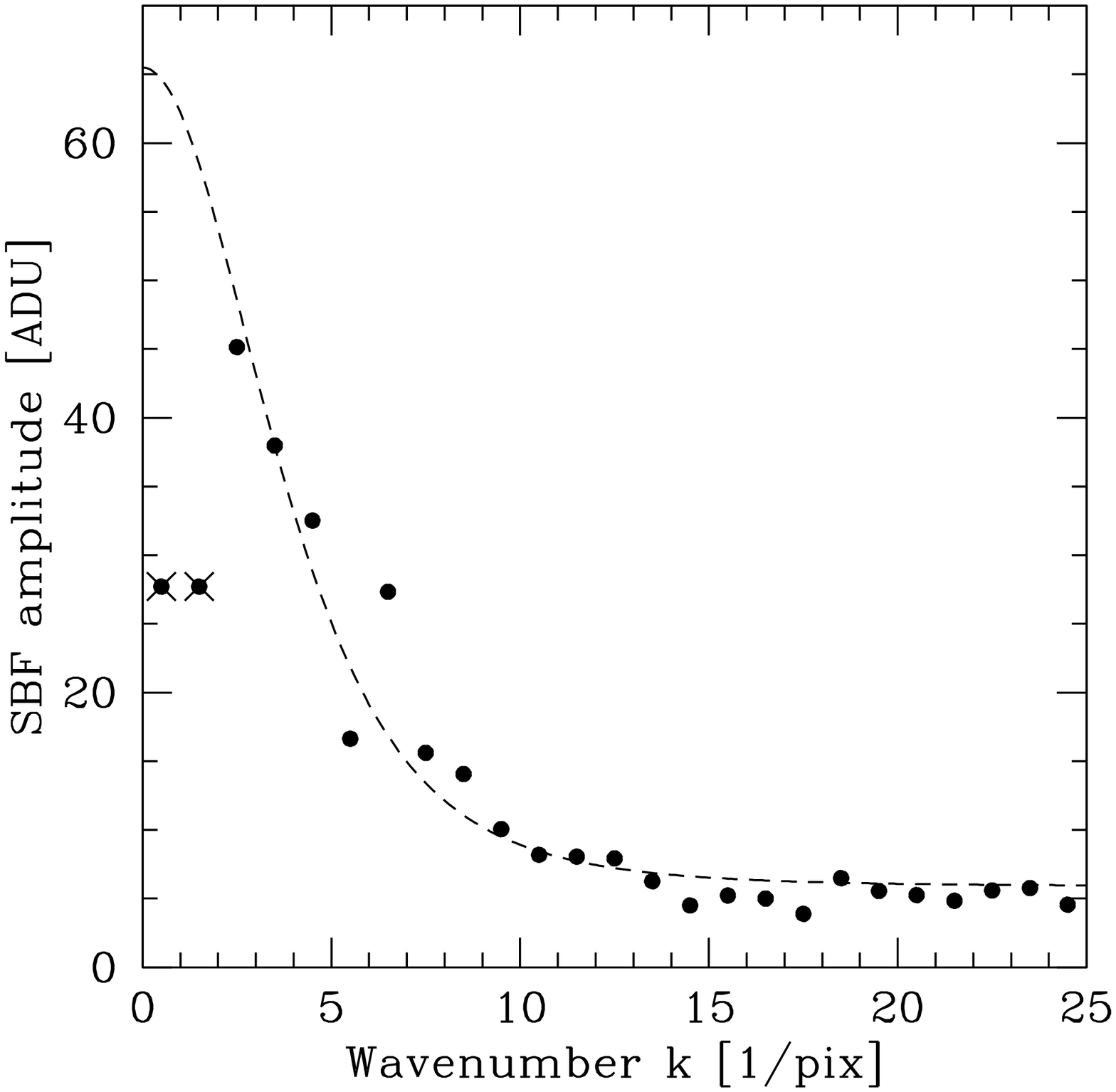,width=4.8cm,height=4.1cm}
\caption{\label{measproc}
Examples of simulated dEs in the $I$ band. For the three upper examples,
the image sequence is: Original image --- model subtracted image ---
Power spectrum of model subtracted, square root divided and cleaned image.
For the lowest example, the sequence is: Model subtracted image --- Cleaned 
model subtracted and square root divided image --- Power spectrum of the latter
image. The different sequence for the lowest image was chosen to illustrate
masking of contaminating point sources.
For the power spectra, the dashed line is the 
best fit of $P(k)=PSF(k)\times P_{\rm 0}+P_{\rm 1}$, when rejecting the points marked 
with crosses. For technical reasons, the SBF amplitude at 0.5 wavenumber is
equal to the amplitude at wavenumber 1.5. For the first example from the top,
the dotted line represents a fit to all datapoints except the outermost and
innermost one. The integration time was 1 hour for all examples.
Parameters of the simulated dEs: 
{\bf 1st panel from top}: $M_{\rm V}=-10.57$ mag, $(m-M)=29.4$ mag, $\mu_{\rm V}$=25.5 
mag arcsec$^{-2}$, 
$(V-I)$=0.88 mag, $\overline{M}_{\rm I}$=-2.4 mag,  
$\delta\overline{m}_{\rm I}=\overline{m}_{\rm I,simulated}-\overline{m}_{\rm I,measured}=-0.22$ mag. 
Seeing 0.5$''$.
{\bf 2nd panel from top}: $M_{\rm V}=-11.13$ mag, $(m-M)=31.4$ mag, $\mu_{\rm V}$=25.1
mag arcsec$^{-2}$, 
$(V-I)$=0.90 mag, $\overline{M}_{\rm I}$=-2.86 mag, $\delta\overline{m}_{\rm I}=0.06$ mag. 
Seeing 0.5$''$.
{\bf 3rd panel from top}: $M_{\rm V}=-13.72$ mag, $(m-M)=31.4$ mag, $\mu_{\rm V}$=23.42
mag arcsec$^{-2}$, 
$(V-I)$=0.99 mag, $\overline{M}_{\rm I}$=-2.44 mag, $\delta\overline{m}_{\rm I}=-0.02$ mag. 
Seeing 1.0$''$.
{\bf 4th panel from top}: $M_{\rm V}=-16.74$ mag, $(m-M)=33.4$ mag, $\mu_{\rm V}$=20.9
mag arcsec$^{-2}$, 
$(V-I)$=1.10 mag, $\overline{M}_{\rm I}$=-1.98 mag, $\delta\overline{m}_{\rm I}=-0.03$ mag. 
Seeing 0.5$''$.}
\end{figure*}
In Fig.~\ref{measproc}, example images and power spectra of simulated dEs are shown.
The innermost pixels which were neglected in the power spectrum fit are especially 
marked. Note the effect of the twice as large seeing of 1.0$''$ in the 
third example: the width of the seeing power spectrum is about half that 
of the other three examples, where the seeing was 0.5$''$.\\
Rejecting the innermost pixels is crucial to determine the correct SBF amplitude,
as low wavenumbers are affected by imperfect galaxy subtraction and large scale
sky gradients in the investigated image. The limiting wavenumber beyond 
which one has to reject pixels has to be determined individually for each galaxy,
as image dimensions and loci and number of contaminating sources change.
We have adopted the following criterion for deciding which pixels to reject or 
not: if the $\chi^2$ of the fit improves by more than a factor of 2 when rejecting
the innermost pixel, it is rejected. Then, the same is tested for the second pixel, and
so on until $\chi^2$ improves by less than a factor of 2. For the examples given in
Fig.~\ref{measproc}, this criterion works fine for the lower three power spectra.\\
Unfortunately, as illustrated in the upper example, things can be more complicated.
If only rejecting the inner- and outermost data point, the obtained fit fits well
to wavenumbers smaller than 6, but underestimates the signal for wavenumbers between
6 and 10 and overestimates the white noise component $P_{\rm 0}$. When rejecting wavenumbers
smaller than 6, the outer part is fit much better.
The difference in $P_{\rm 1}$ between the two possibilities is considerable, about 40\%.
In cases like that, the uncertainty in which pixels to reject or not is the major source
of error. As is noted in Fig.~\ref{measproc}'s caption, the difference $\delta\overline{m}_{\rm I}$
between simulated and measured $\overline{m}_{\rm I}$ obtained when
rejecting the inner 5 pixels is much smaller than for only rejecting the innermost pixel. 
Therefore, whenever fits to the outer and inner part of the power spectrum differed 
considerably, more emphasis was put on fitting well the outer part.
\\ 
However, further out 
in the FT profile the white noise component starts to dominate over the PSF FT. 
As our simulations were performed on artificial {\bf dwarf} galaxies, the image 
portions chosen for the SBF measurements had relatively small dimensions 
of typically 30 to 60 pixel (6 to 12~$''$). 
Therefore the wavenumber range over which the 
amplitude of the PSF is determined is only of the order of 10 or fewer
independent data points. This small number, together with the uncertainty in which wavenumbers
to disregard or not, is the major source of uncertainty for the SBF measurements
we performed. For the simulated galaxies at 1.0$''$ seeing, the wavenumber range
over which to perform the fit to the PS is only half of that for 0.5$''$ seeing,
which means the uncertainty at 1.0$''$ seeing is significantly higher than for 0.5$''$ 
seeing.\\
\begin{figure}
\begin{center}
\psfig{figure=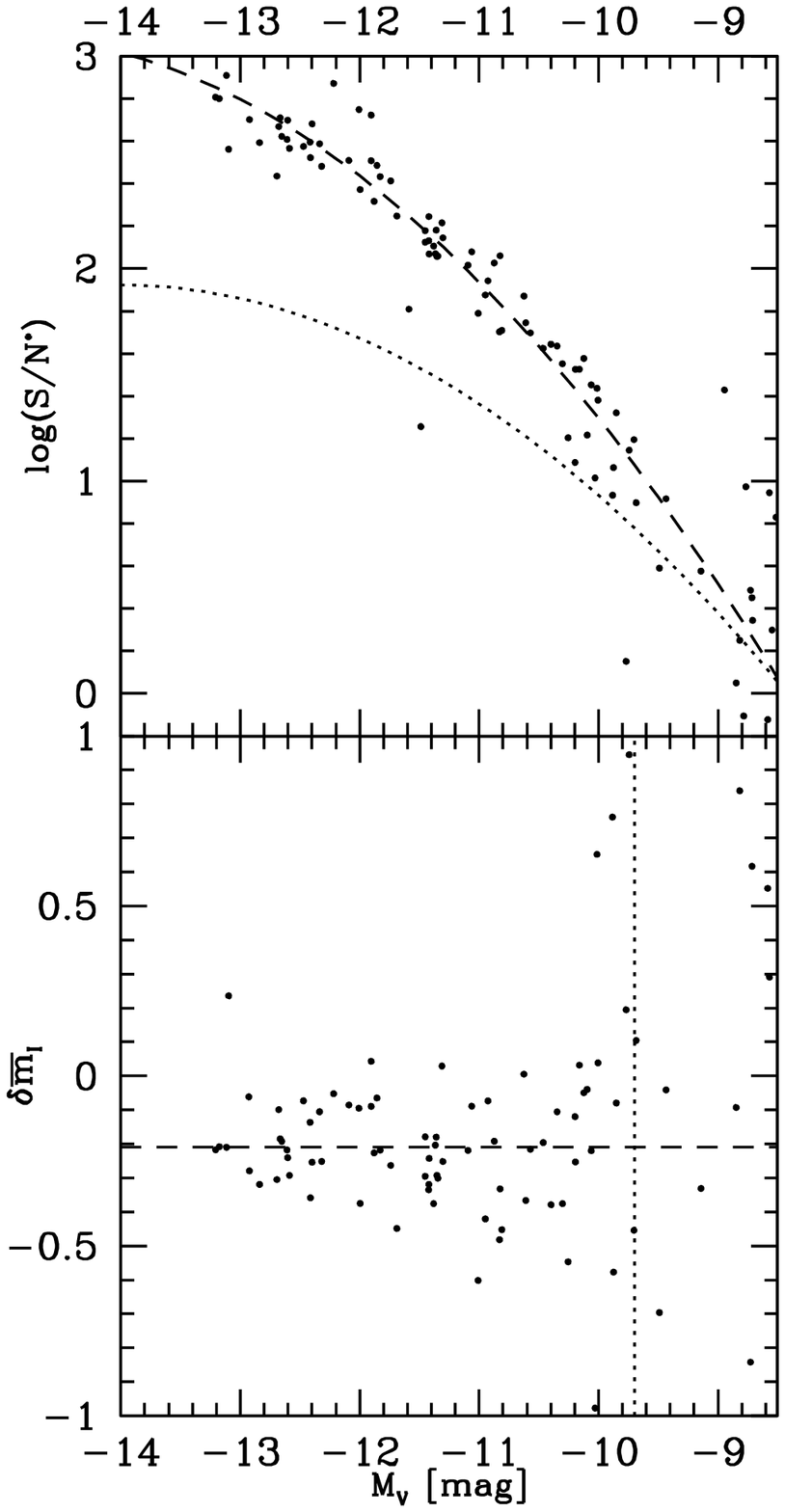,width=4.3cm}
\psfig{figure=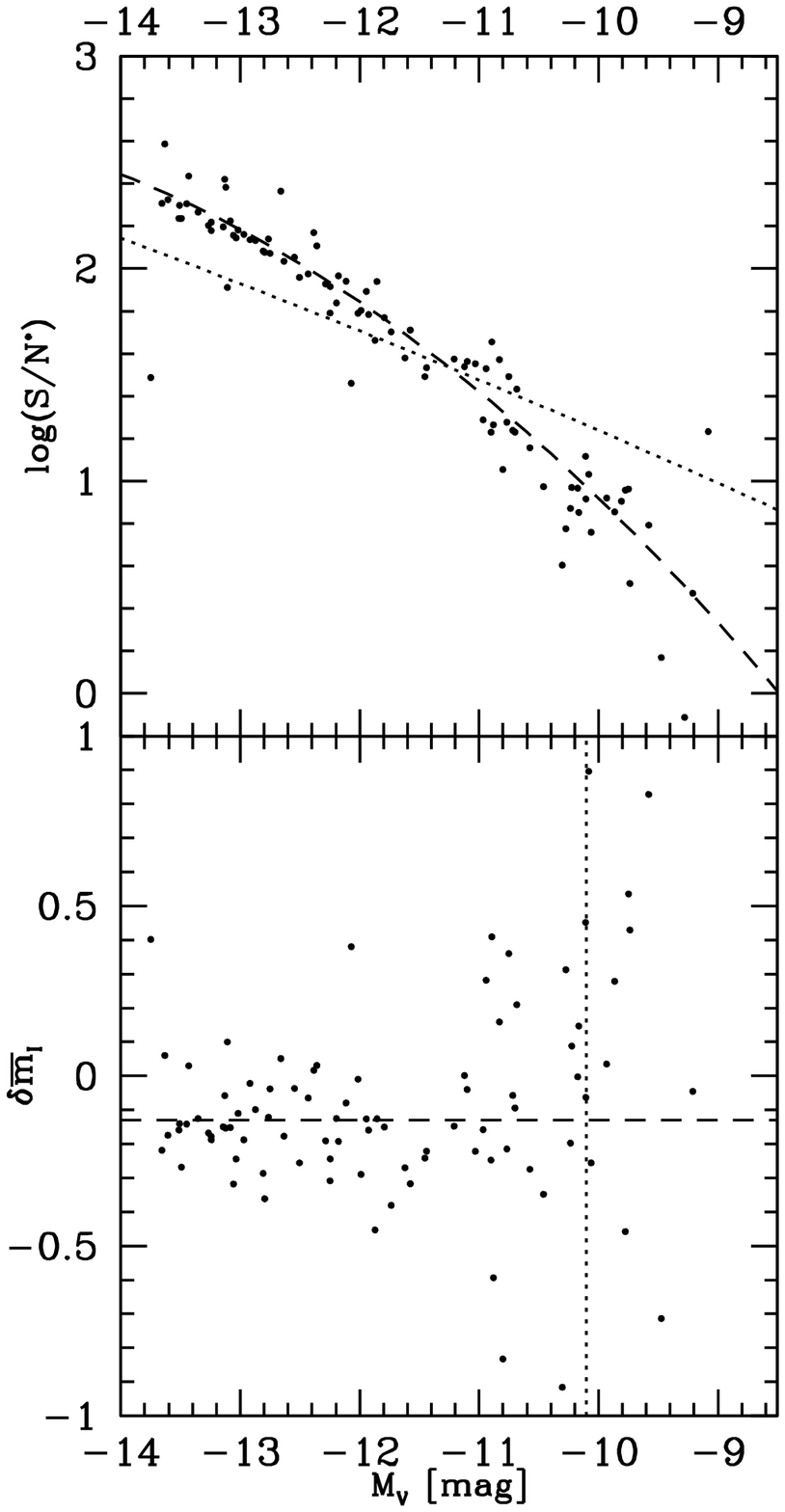,width=4.3cm}
\end{center}
\caption[]{\label{resultsseeing294}Results of the simulations for 
$(m-M)$=29.4 mag and zero point $ZP=27.0 + 2.5*log(3600)$ mag. 
{\bf Left panel}: Seeing=0.5$''$. {\bf Right panel}: Seeing=1.0$''$. {\bf Top panel}:
Logarithm of $S/N^*$ (definition see text) plotted vs $M_{\rm V}$. The dashed
line is a 2nd order polynomial fit to the data points. The dotted line
is the fit corresponding to the canonically defined S/N (see text).
{\bf Bottom panel}: 
$\delta\overline{m}_{\rm I}=\overline{m}_{\rm I,simulated}-\overline{m}_{\rm I,measured}$ 
plotted vs. $M_{\rm V}$. For magnitudes fainter than the limiting magnitude $M_{\rm V}^*$ indicated 
by the dotted vertical line, 
more than 50\% of the measured galaxies have $S/N^*<6$ or a deviation
of more than 0.5 mag from the horizontal dashed line. The latter one denotes the mean 
$\overline{\delta\overline{m}_{\rm I}}$ for the measured galaxies brighter than $M_{\rm V}^*$.
$\overline{\delta\overline{m}_{\rm I}}$ and $M_{\rm V}^*$ were determined iteratively.}
\end{figure}
\begin{figure}
\begin{center}
\psfig{figure=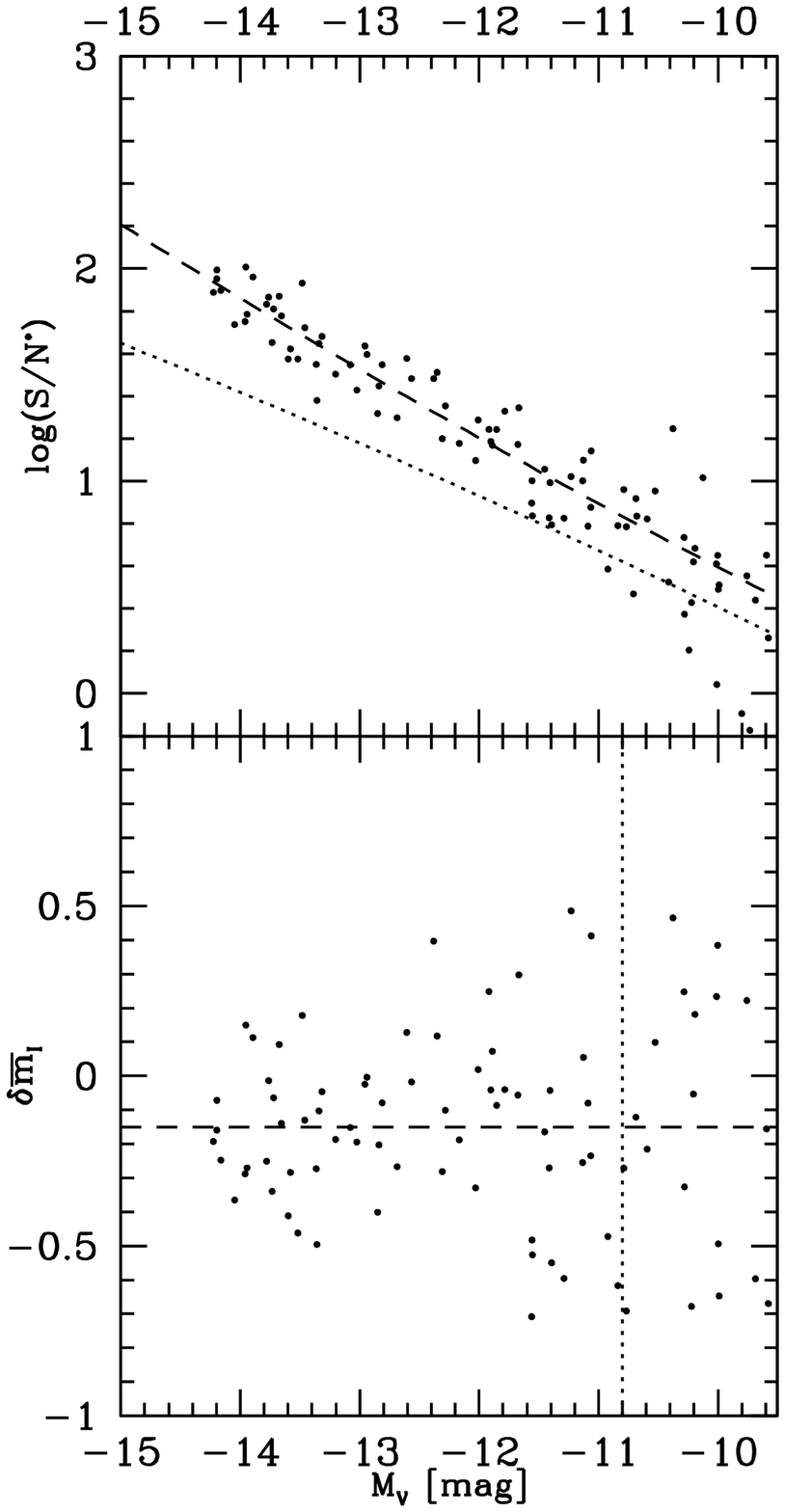,width=4.3cm}
\psfig{figure=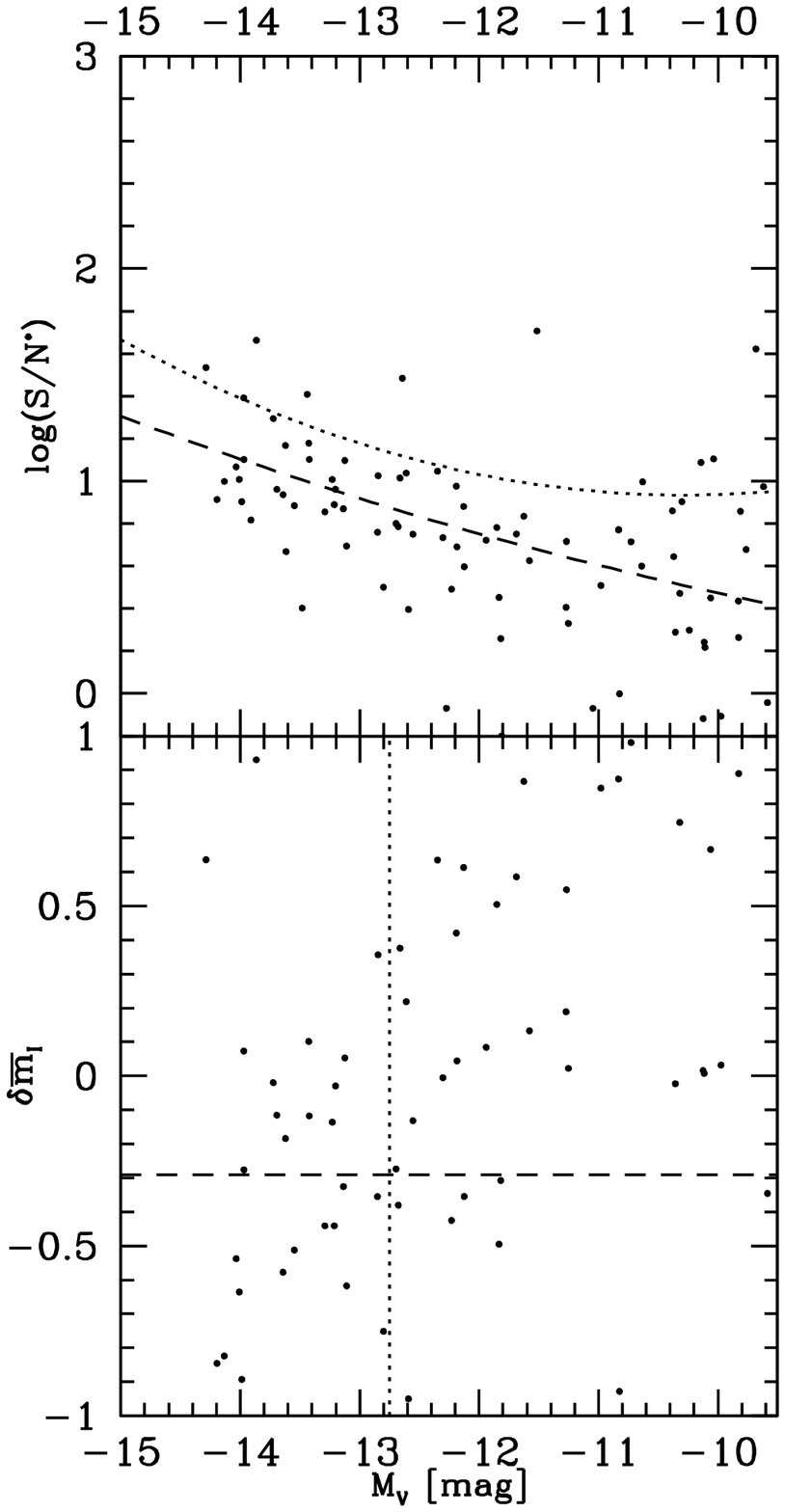,width=4.3cm}
\end{center}
\caption[]{\label{resultsseeing314}Results of the simulations for 
$(m-M)$=31.4 mag. Symbols/lines as in Fig.~\ref{resultsseeing294}.}
\end{figure}

\begin{figure}
\begin{center}
\psfig{figure=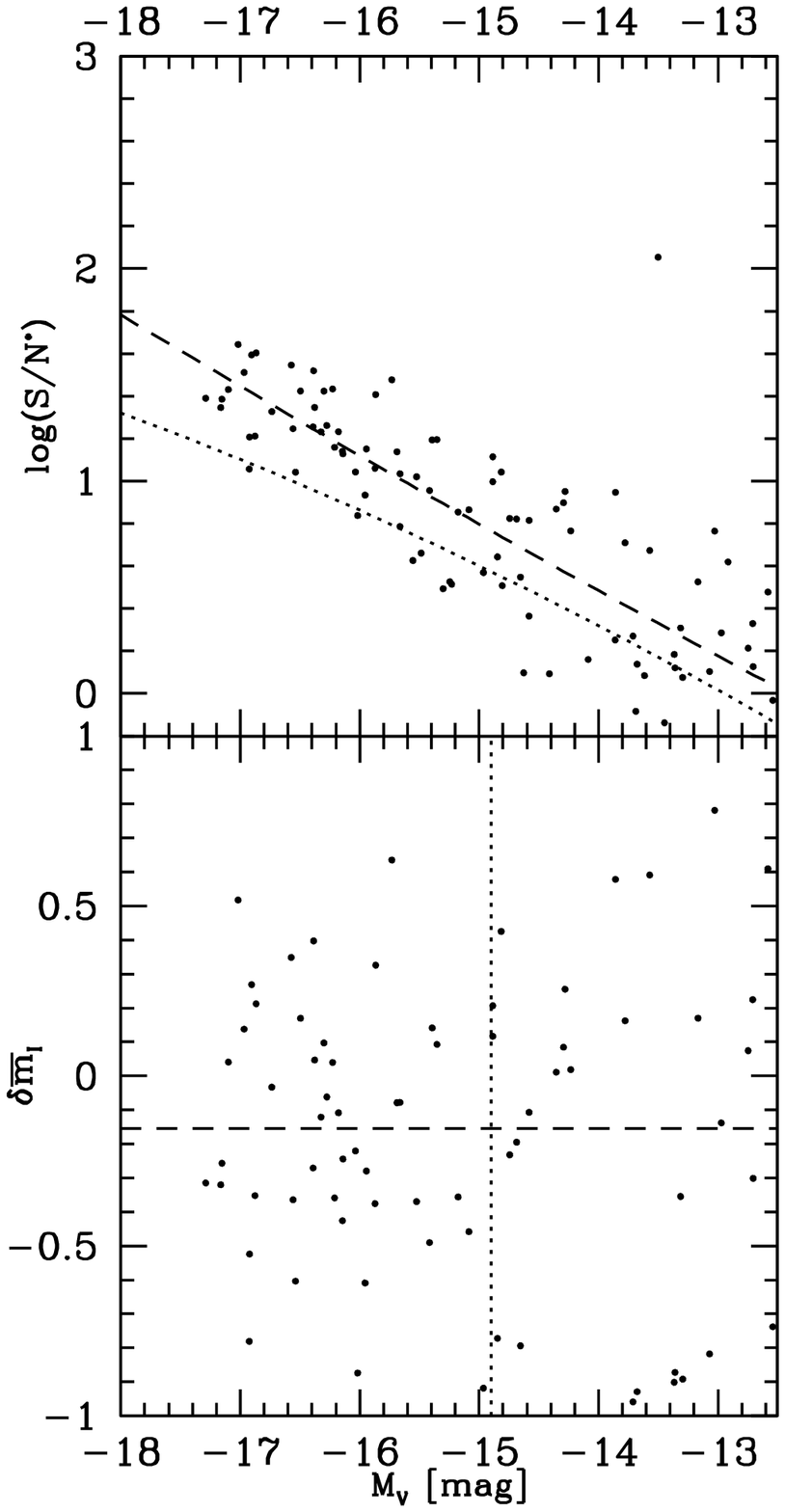,width=4.3cm}
\psfig{figure=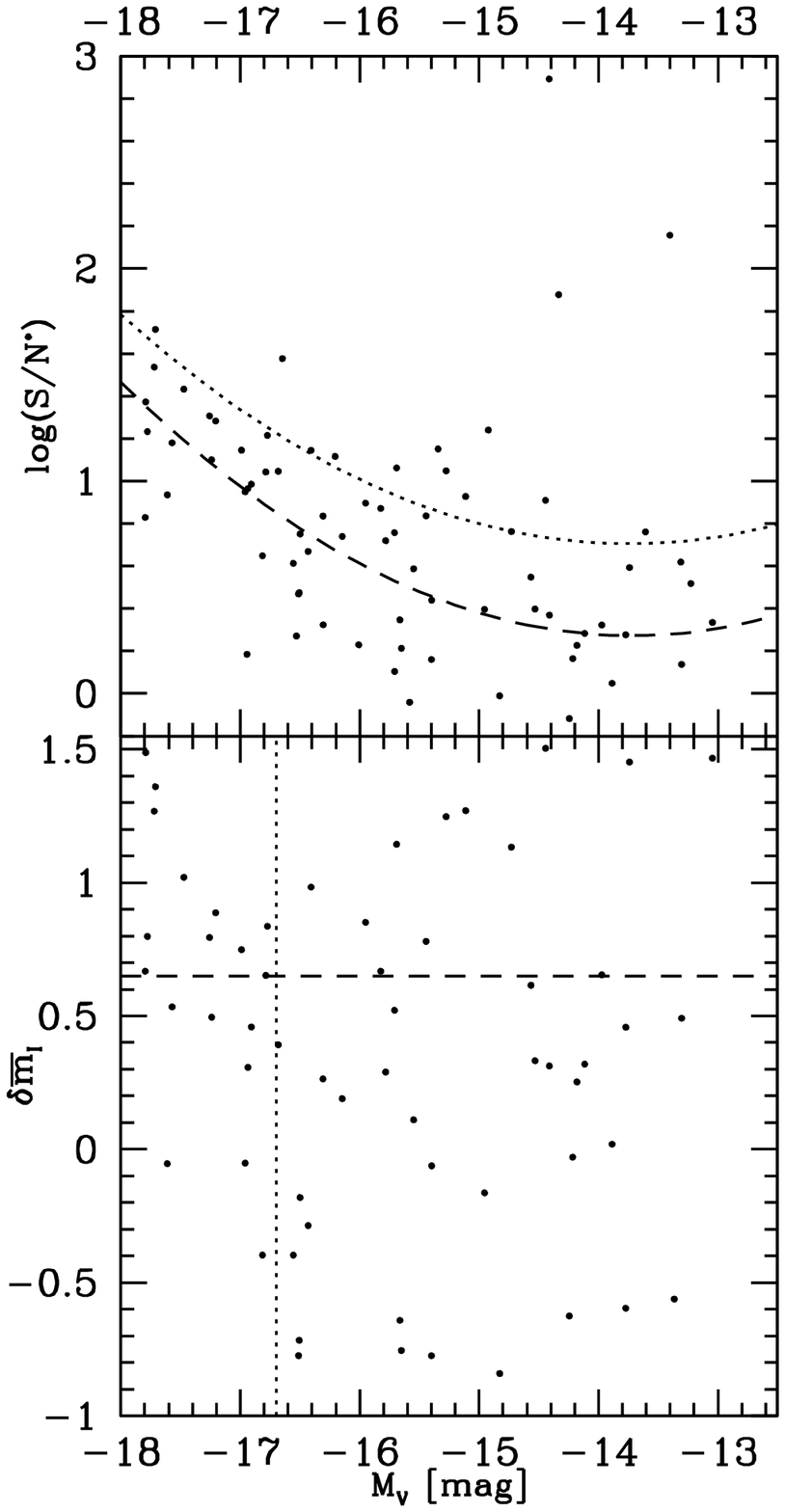,width=4.3cm}
\end{center}
\caption[]{\label{resultsseeing334}Results of the simulations for 
$(m-M)$=33.4 mag. Symbols/lines as in Fig.~\ref{resultsseeing294}.}
\end{figure}

\begin{figure*}
\begin{center}
\psfig{figure=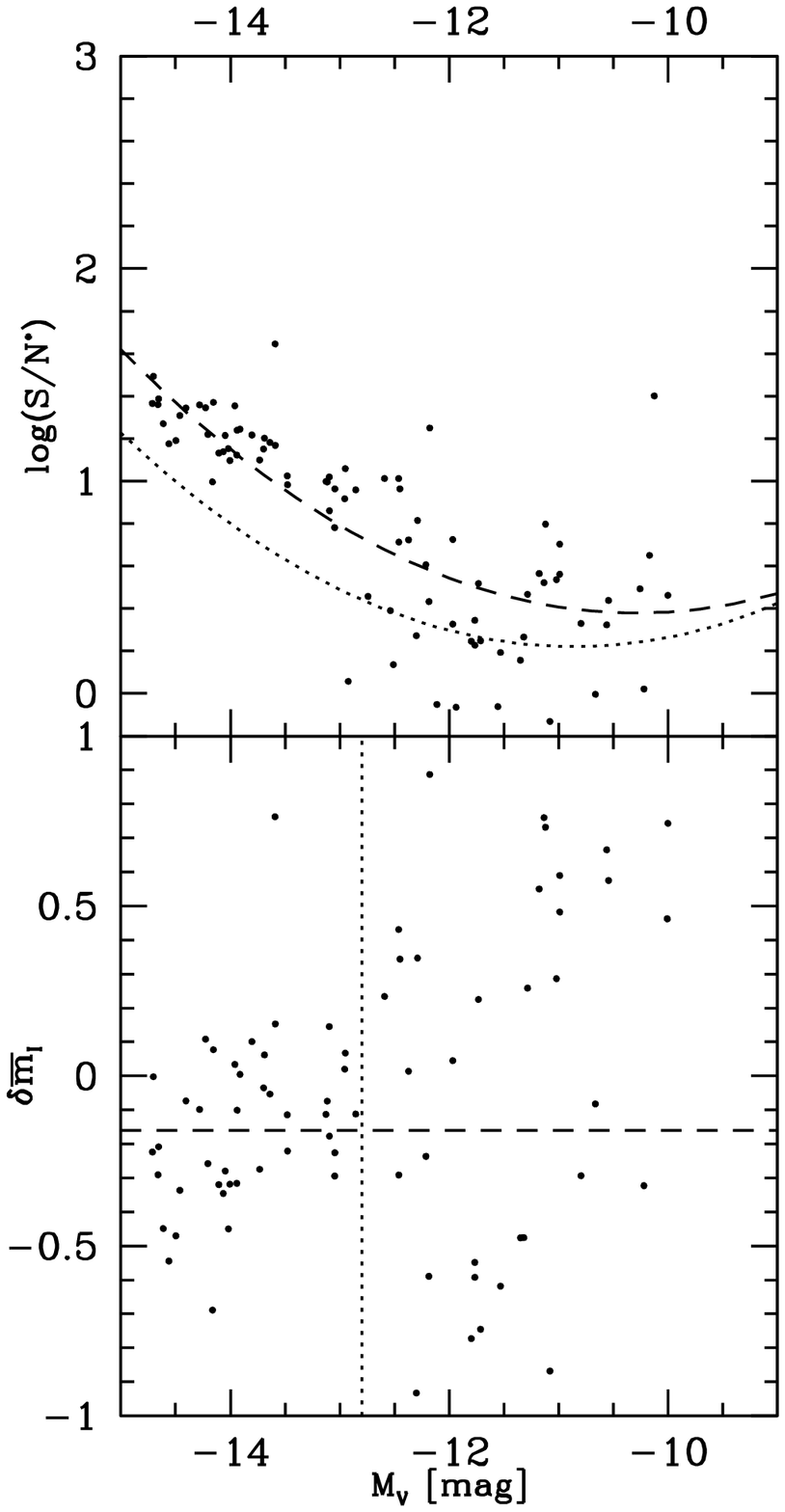,width=4.cm}
\psfig{figure=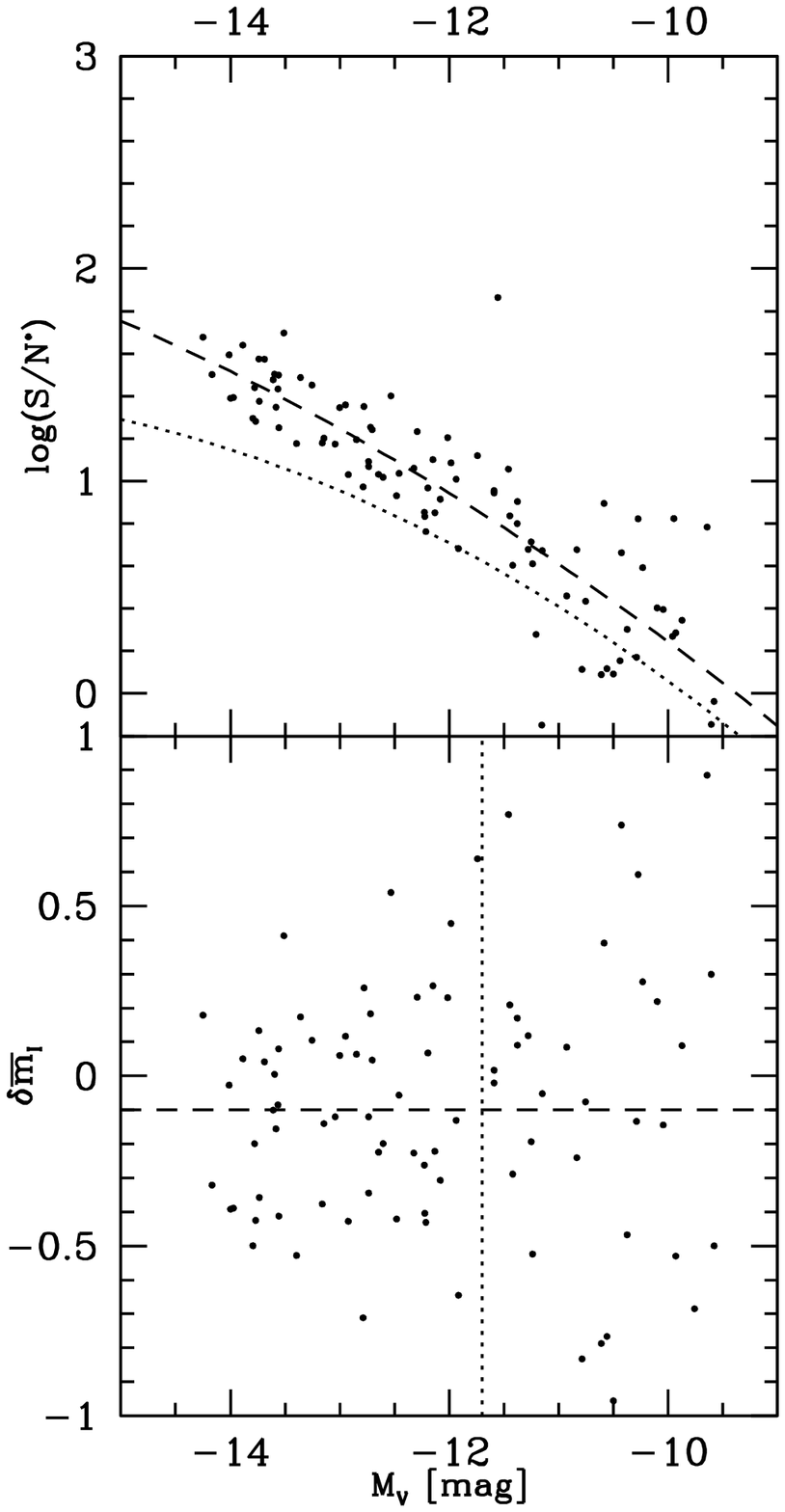,width=4.cm}
\psfig{figure=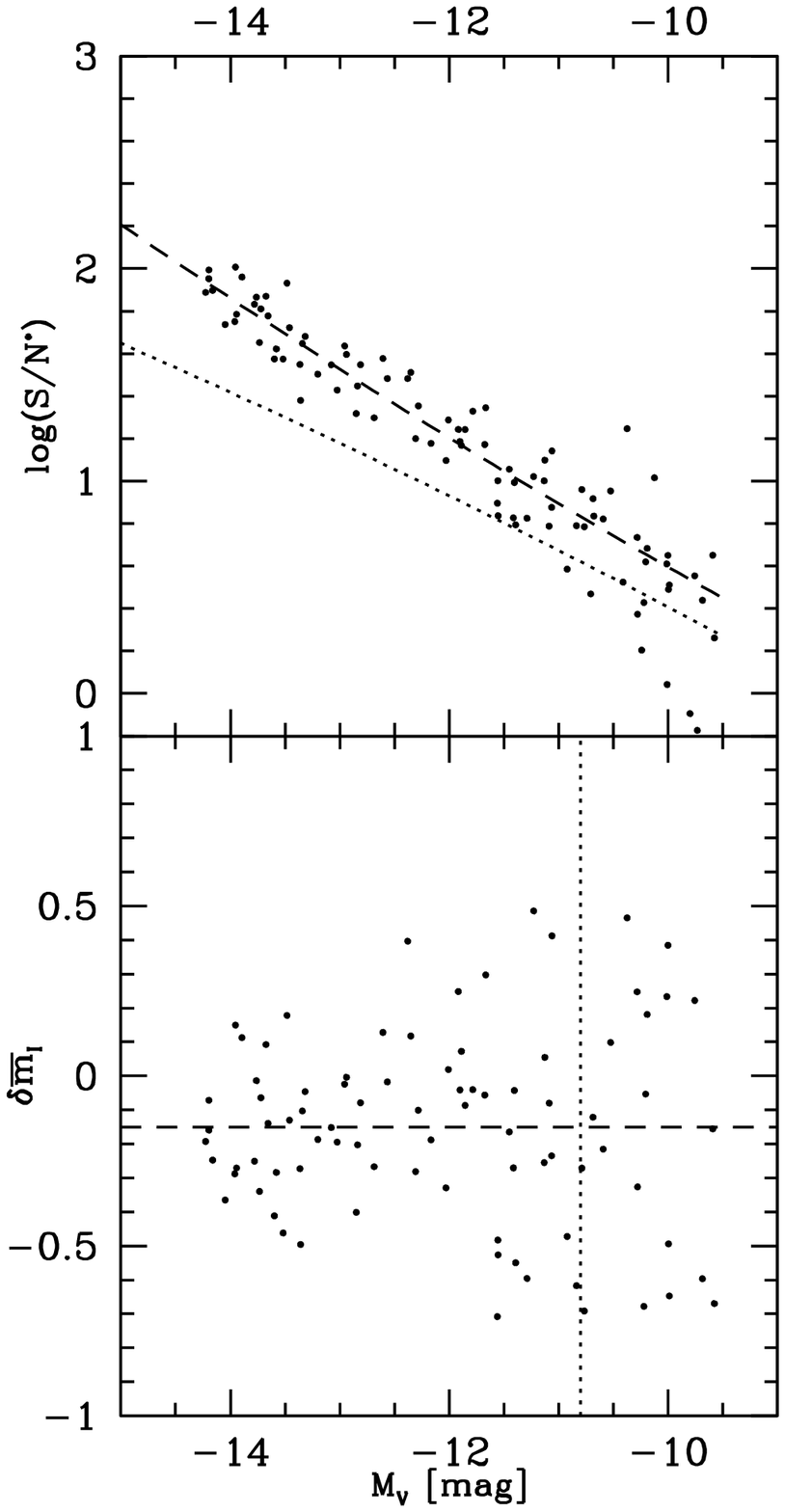,width=4.cm}
\psfig{figure=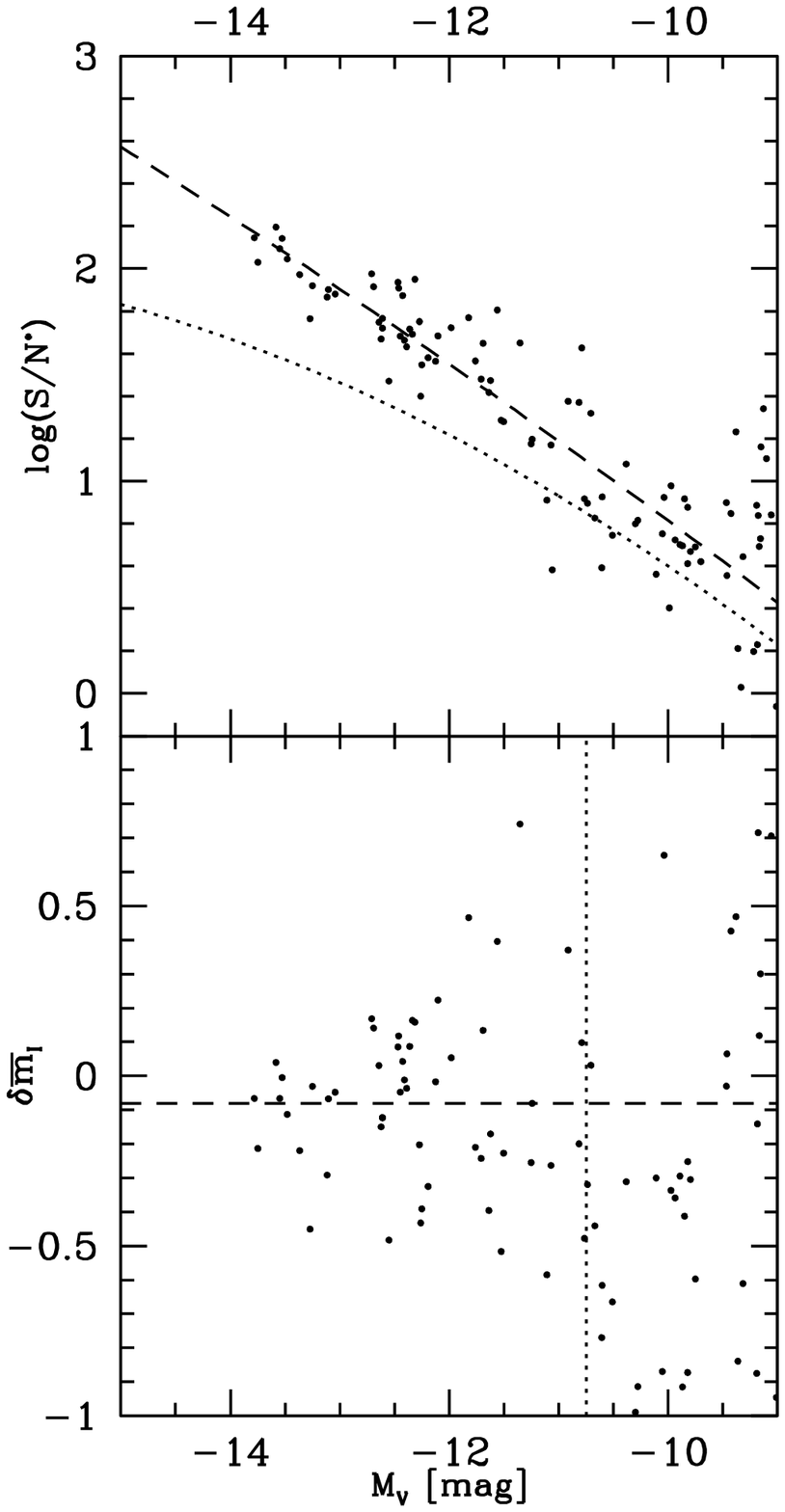,width=4.cm}
\end{center}
\vspace{-0.2cm}
\caption[]{\label{results31405}Results of the simulations for $(m-M)$=31.4 and
seeing=0.5$''$ at varying integration time / zero point ZP=27.0+2.5*log(t).
From left to right: t= 900; 1800; 3600; 7200 seconds.}
\end{figure*}
\section{Results of the simulations}
\label{simres}

In Figs.~\ref{resultsseeing294} to~\ref{results31405}, the results of the 
simulations are shown. Figs.~\ref{resultsseeing294} to~\ref{resultsseeing334} 
show the results for the three different distance moduli with 3600 seconds integration
time: 
$S/N^*$ and $\delta\overline{m}_{\rm I}$ are plotted vs. $M_{\rm V}$ for the two different seeing
values. In Fig.~\ref{results31405}, the same observables are plotted for four different
integration times at a fixed 31.4 mag distance 
modulus and 0.5$''$ seeing.\\
For the reasons mentioned in Sect.~\ref{meassbf} we preferred to plot the modified signal
to noise $S/N^*$ vs. $M_{\rm V}$ instead of the canonical S/N, as $S/N^*$  takes into account 
seeing effects and thus 
allows a comparison
between measurements obtained with different seeing. A 2nd order polynomial fit 
to $S/N^*$ vs $M_{\rm V}$ is plotted
over the data points. The corresponding fit for the canonical S/N is given as well.
It is interesting to note that for 0.5$''$ seeing the modified $S/N^*$ is consistently
higher than the canonical S/N, while for 1.0$''$ it is the opposite. This shows that for
our simulations, 0.75$''$
is about the limiting seeing below/above which $S/N^*$ becomes higher/lower than the 
canonical S/N.\\
One can see that for the highest $S/N^*$ data, the relative accuracy of the SBF method 
is of the order of 0.1-0.2 mag, i.e. 5-10\% in distance.
By glancing at the bottom panels of 
Figures~\ref{resultsseeing294}~to~\ref{results31405}, one immediately notices 
an average offset between
simulated and measured SBF amplitude of the order of 0.15 mag in the sense of measuring
too faint SBF. The mean offset, when disregarding
the two most extreme offsets, is 0.14 mag. This offset is dealt with in more detail
in Sect.~\ref{bias}.\\
\subsection{Limiting absolute magnitudes}
\label{limmag}
For each of the 9
simulated sets of dEs a limiting absolute magnitude $M_{\rm V,lim}$ was determined, below 
which the determination 
of cluster membership is not reliable anymore. These are the two conditions for
reliable cluster membership determination we adopted:\\ 
First, the difference between offset corrected 
measured and simulated $\overline{m}_{\rm I}$ must be smaller than 0.5 mag. The mean offset 
with regard to which the measurement difference is defined
is denoted as $\overline{\delta\overline{m}_{\rm I}}$ and indicated in Figures~\ref{resultsseeing294} 
to~\ref{results31405}. The value of 0.5 mag
was chosen as it is about equal to the maximum uncertainty in deriving $\overline{M}_{\rm I}$ from
$(V-I)$, see Sect.~\ref{degeneracy}. To require a higher measurement accuracy than the intrinsic
uncertainty of the method would be unnecessary. \\
Second, the modified $S/N^*$ of the
measurement must be higher than 6. 
This limit was adopted to avoid an SBF measurement of an object with $S/N^*<4$ mimicking
 a $S/N^*$ of the order of 5 or 6 because of
a measured $\overline{m}_{\rm I}$ 0.5 mag brighter than simulated.\\$M_{\rm V,lim}$ is then defined as the 
absolute magnitude at which 50\% of the measured galaxies fulfil the above criteria. It
is indicated for each set in Figures~\ref{resultsseeing294} 
to~\ref{results31405}. As $\overline{\delta\overline{m}_{\rm I}}$ and $M_{\rm V,lim}$ depend on each 
other, they were determined iteratively.\\
In Table~\ref{simrestable}, $M_{\rm V,lim}$ is tabulated. One can see that for the two smaller
distance moduli 29.4 and 31.4, the SBF-Method can reach very faint magnitudes. For galaxies 
with $M_{\rm V}\simeq -10$ to $-11$ mag within a distance of about 20 Mpc, reliable SBF
measurements with accuracies better than 0.5 mag can be obtained at about 0.5$''$ seeing, 
the given zeropoint of 27 mag and an integration time of 1hr. The SBF-Method is therefore
a very valuable tool for extragalactic distance measurements, even for the faintest galaxies.\\
What are the effects of
varying seeing, integration time and distance modulus?\\
\begin{table}
\begin{center}
\begin{tabular}{l||ll}
\small
 & 0.5$''$ & 1.0$''$\\\hline\hline
29.4&-9.7&-10.1\\
31.4&{\it -12.8}; {\it -11.7}; -10.8; {\it -10.8}&-12.75\\
33.4&-14.9&-16.75\\
\end{tabular}
\end{center}
\caption[]{\label{simrestable}Limiting absolute magnitude $M_{\rm V,lim}$ for the different
combinations of distance modulus (denoted in the left column) and seeing (denoted in
the top line). The integration time $t$ was 3600 seconds except for the four different 
values at 0.5$''$ and 31.4 which correspond to four different
integration times $t$=900, 1800, 3600 and 7200 seconds from left to right. The zero point
including integration time $t$ is given by $27 + 2.5*log(t)$.}
\normalsize
\end{table}
\subsubsection{Effects of varying seeing}
Table~\ref{simrestable} shows that for the distance moduli 31.4
and 33.4, increasing the seeing by a factor of 2 brightens $M_{\rm V,lim}$ by about
2 mag. This corresponds to about 1.4 mag in central surface brightness $\mu_0({\rm V})$, or a
factor of 3.6 in central intensity. This increase by almost a factor of 4 in central intensity
is plausible: the pixel-to-pixel fluctuations are smoothed by a
factor of 2 at 1.0$''$ seeing compared to 0.5$''$ while the SBF amplitude is 
proportional to the square root of the intensity; therefore the intensity must be increased 
by a factor of 4 to compensate for the smoothing caused by the twice as large seeing.\\
The ``rule'' extracted from that behaviour is: Increasing seeing by a factor $x$ needs increase
of intensity by a factor of $x^2$ to be compensated.\\
For the 29.4 mag distance modulus, there is no big difference between 0.5$''$ and 1.0$''$ seeing.
This is because at the faint magnitudes around $-10$ mag, the central surface brightness is about
25 mag arcsec$^{-2}$ in $I$, which is only 2-3 $\sigma$ above the sky standard deviation, i.e. 
the mean surface brightness is close to the detection limit, and measuring its 
fluctuations is very difficult, even if the SBF amplitude is not much smaller than the mean
surface brightness. That is why going from 1.0$''$ to 0.5$''$ seeing, no significant
improvement of limiting absolute magnitude is reached for the 29.4 mag distance modulus.\\
\subsubsection{Effects of varying integration time / zero point}
What is the necessary scaling in integration time $t$ to account for varying seeing? At 31.4 mag
distance modulus, only $t$=900 seconds are needed for 0.5$''$ seeing to
reach the same $M_{\rm V,lim}$ as for 1.0$''$ seeing and $t$=3600 seconds. \\Thus, increasing 
seeing by a factor $x$ needs increase of integration time by a factor of $x^2$ to be compensated.
This is equivalent to keeping integration time fixed and increasing the zero point by
2.5*log($x^2$).\\
Table~\ref{simrestable} shows that increasing the integration time $t$ by a factor of 
2 results in a 1 magnitude fainter $M_{\rm V,lim}$, 
or about a factor of 2 in central intensity. As S/N and SBF amplitude are proportional to 
$\sqrt{t}$ and $\sqrt{I}$, respectively, this result should be expected.\\
Thus, increasing integration time $t$ by a factor $x$
allows SBF measurement for objects with
central intensity fainter by the same factor $x$. This is equivalent to keeping integration 
time fixed and increasing the zero point by 2.5*log($x^2$).\\
No notable change in
limiting magnitude is seen when increasing $t$ from 3600 to 7200 seconds. This might be partially
due to statistical reasons, but the major reason is that the mean surface brightness is close to
the detection limit (about 5 sigma above the sky noise) and the angular
extent of the simulated dEs is only a few arcseconds. For galaxies with $M_{\rm V}\simeq-11$ mag,
the additionally detected region when going from 3600 to 7200 seconds carries no measurable
SBF signal anymore.\\
\subsubsection{Effects of varying distance modulus}
The strength of the SBF relative to the underlying
mean surface brightness decreases linearly with distance. As the SBF are proportional to the square
root of the intensity, the intensity must increase by a factor of $x^2$ when distance increases
by a factor of $x$ to compensate for that. Table~\ref{simrestable} shows that 
increasing distance modulus by 2 mag results in a 3-4 mag brighter limiting magnitude $M_{\rm V,lim}$.
This corresponds to about a factor of 10 in central intensity, which is slightly more
than the expected value of 2.5$^2$=6.25. The reason for this is that the angular area over which
the SBF signal is sampled is smaller at 2.5 higher distance for the same object.\\
\subsubsection{A simple rule}
Summarizing the scaling relations found in the last three subsections, we give the following
rule to calculate the limiting magnitude $M_{V\rm new}^*$ at new observing conditions 
different to the
reference ones adopted in our simulations:\\
\begin{equation}
\label{rule}
\small
M_{V\rm new}^*=M_{V\rm ref}^*-6.64*log(\frac{s_{\rm new}}{s_{\rm ref}})-1.33*(ZP_{\rm ref}-ZP_{\rm new})
\small\normalsize
\end{equation}
$M_{V\rm ref}^*$ is the limiting magnitude calculated from our simulations at the given distance
modulus. $s_{\rm ref}$ is the seeing diameter in our simulations, $s_{\rm new}$ the new seeing
diameter. $ZP_{\rm ref}$ is the total zero point in the simulations, i.e. 
$ZP_{\rm ref}=27.0+2.5*log(t)$ with $t$ being the total integration time. $ZP_{\rm new}$ is then
the new total zero point. This all refers to a gain of 1, i.e. the zero point is expressend
in terms of electrons and not ADU.\\
Note, however, that equation~\ref{rule} is restricted to cases where the mean surface brightness 
of the galaxy is significantly higher than the sky noise. As when the surface brightness gets 
too close to 
the sky noise (less than about 5 sigma, see the former three subsections), changing integration
time or seeing does not have strong effects on SBF detectability.\\
\subsection{A bias in the simulations}
\label{bias}
To find out the reasons for the 0.14 mag average offset between simulated and measured 
$\overline{m}_{\rm I}$, a number of tests were performed:\\
First, an area of constant surface brightness and SBF amplitude was simulated, but not convolved 
with the seeing. After subtracting the mean brightness, 
dividing by its square root and calculating the power spectrum, the resulting image should have a
mean equal to the SBF amplitude. This was the case to within 1\%, independent of surface
brightness and SBF amplitude.\\
Second, an area of intensity zero was simulated with only one pixel given an intensity $I_0$
different from zero. This area was then convolved with the seeing profile. The mean of the 
resulting image was $0.99\times I_0$. I.e. by restricting the seeing psf simulation to 7 times 
the FWHM, about 1\% flux is lost. This is because we chose a realistic Moffat profile, 
which has stronger wings than the commonly used Gaussian.\\
The two above tests show that the algorithm used to implement and measure pixel-to-pixel
SBF works correctly, and that
convolution with the seeing profile to 7 times the FWHM imposes a negligibly small flux loss.\\
The final check performed was to simulate areas of several hundred pixel side length with
constant surface brightness and SBF amplitude, which were then convolved with the seeing.
After subtraction of the mean brightness, 
division by its square root, calculation of the power spectrum and azimuthally averaging, 
the SBF amplitude according to equation~(\ref{azimut})
-- i.e. in the limit of k=0 -- 
was too faint by 0.10 to 0.15 mag, independent of the strength of the simulated SBF, and for both 
seeing values 0.5 and 1.0$''$. This offset corresponds to the one found in the 
main simulations. It shows that
recovering the underlying pixel-to-pixel fluctuations from seeing convolved images is 
subject to small, but non-negligible loss of fluctuation signal, at least in our simulations.
\\
Already Tonry \& Schneider (\cite{Tonry88}) noted a bias towards measuring too faint SBF of the 
order of 10\% in distance or 0.2 mag in distance modulus, and attributed this to truncation
of the seeing profile in their simulations. As mentioned above, in our case seeing truncation
imposes a negligible flux loss.\\
Zero point problems of the order of 0.15 mag are certainly a matter of concern when one aims at
determining absolute values like $H_{\rm 0}$. However, our aim was to determine limiting magnitudes
and surface brightnesses for cluster membership determination of dEs when using the SBF-Method.
These measurements all rely on relative distances and are therefore independent of any zero point
offset. The offset found in our simulations does not significantly change the statements made in 
the former sections.\\
We note that unresolved background galaxies generally increase the measured SBF signal. 
However, for our simulations this contribution is negligible. Using formula (13) of 
Jensen et al. (\cite{Jensen98}) for the $I$-band and inserting the values
used for the magnitude distribution of background objects, we get for the relative contribution 
of background galaxies to the SBF signal at 33.4 distance modulus a value 
of the order of 2-4\%, depending on seeing and the galaxy's magnitude. For the distance moduli
29.4 and 31.4, the contribution is below 1\%.\\
\subsection{Comparing real SBF data with simulations}
\label{applcen}
\begin{figure}
\begin{center}
\psfig{figure=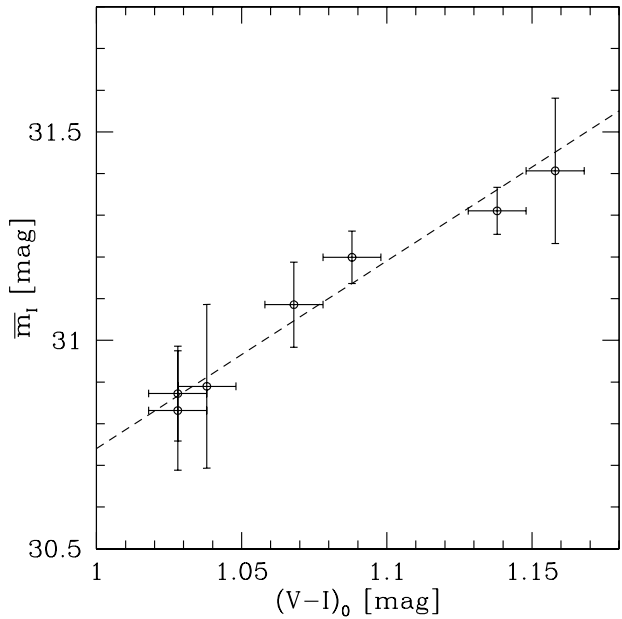,width=8.6cm}
\end{center}
\vspace{-0.2cm}
\caption[]{\label{cencolsbf}Apparent SBF magnitude $\overline{m}_{\rm I}$ plotted vs. 
$(V-I)_0$ for 6 Centaurus cluster 
dEs (from Mieske et al. \cite{Mieske03}). The dashed line is a fit to the 
data when adopting a slope of 4.5, as 
in Tonry's equation~(\ref{sbfrel}). The corresponding distance modulus is 33.15 
$\pm$ 0.04 mag. 
When fitting slope and zero point at the same time, the resulting slope differs by 
less than one $\sigma$ from 4.5.}
\end{figure}
\begin{figure}
\begin{center}
\psfig{figure=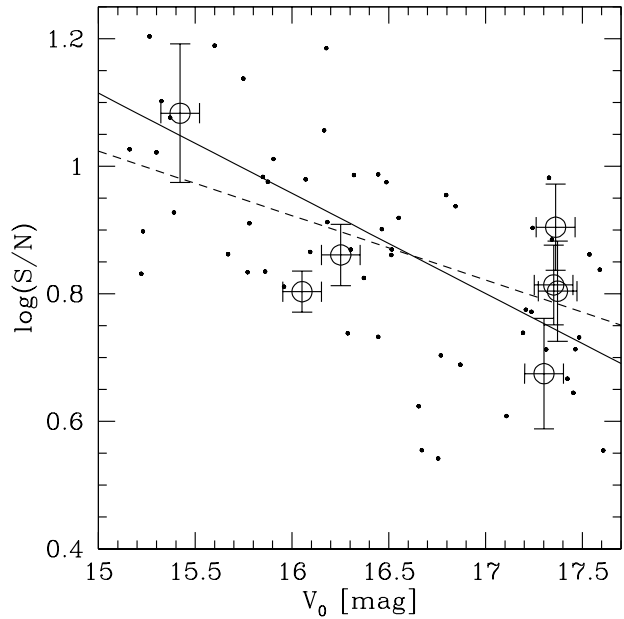,width=8.6cm}
\end{center}
\vspace{-0.2cm}
\caption[]{\label{sbfcensim} 
S/N of the SBF measurement vs. apparent $V_0$. 
Large circles are the real Centaurus data (from Mieske et al. \cite{Mieske03}). 
Dots represent galaxies simulated according to the Centaurus data's colour-SBF relation, 
colour-magnitude relation and surface brightness-magnitude relation. 
The dashed 
line represents a fit to the real data, the solid line is a fit to the simulated data.
Both fits are consistent with each other within the error ranges.} 
\end{figure}

The S/N achieved in SBF measurements for 6 bright Centaurus 
cluster dEs from VLT FORS1 images is compared with the S/N obtained from simulations 
tuned to reproduce the measured values.\\
Fig.~\ref{cencolsbf} shows the colour-SBF relation found for the 6 galaxies (Mieske et al., 
\cite{Mieske03}). They cover a magnitude range of $15.4<V_0<17.4$ mag. 
The colour-SBF relation from Fig.~\ref{cencolsbf}, a colour-magnitude relation 
and surface brightness-magnitude relation were fit to the measured values of the 
Centaurus dEs, and 64 galaxies were simulated according to these relations. Their SBF 
amplitude was measured as described in Sect.~\ref{simexpl}. In Fig.~\ref{sbfcensim}, 
the log(S/N) values of the real measurements
are plotted over the results for the simulations. A line is fit to both real and 
simulated data.
One can see that the simulations do not overestimate the S/N of the real data. 
Both fits are consistent with each other.\\
This consistency between real and simulated data confirms that the simulations 
presented in the previous sections are a good approximation of reality. The applied
generalizations like zero ellipticity and purely exponential profile apparently do not
introduce a notable bias towards too high or too low S/N.\\
\section{Summary and conclusions}
Extensive simulations of SBF measurements on dEs for three different distance
moduli 29.4, 31.4 and 33.4 mag, two different seeings 0.5$''$ and 1.0$''$ and four 
different observing times 900, 1800, 3600 and 7200 seconds have been presented.
For each of the simulated sets of dEs, the limiting magnitude $M_{\rm V,lim}$ below which
a distance measurement is not reliable anymore has been determined.
It was shown that for distances $\le$ 20 Mpc, the SBF method can yield reliable
cluster membership of dEs down to very faint limiting magnitudes, 
e.g. $M_{\rm V,lim}\simeq -10$ mag for
a distance of 7.5 Mpc, and $M_{\rm V,lim}\simeq -11$ mag for 19 Mpc distance, at 1hr integration
time, 0.5$''$ seeing and a zero point of 27 mag in the $I$-band. For the SBF measurements,
a modified signal to noise $S/N^*$ has been defined that incorporates the seeing 
dependence of SBF detectability.\\
The effects of varying seeing, integration time and distance modulus on the limiting 
absolute magnitude are investigated. A number of simple rules, confirmed by theoretical 
considerations, are derived in order to calculate limiting magnitudes, needed 
integration times or seeing for observing conditions different to the ones
adopted for our simulations.\\
It is pointed out that the total uncertainty in obtaining a distance modulus for a dE 
with SBF measurements 
is the quadratic sum of the measurement uncertainties for $\overline{m}_{\rm I}$
 and the uncertainties in deriving $\overline{M}_{\rm I}$ from $(V-I)$. As both uncertainties
are 0.5 mag or smaller, the worst possible measurement accuracy is of the order of 0.65 mag 
or 35\% in distance. This would apply to very faint and blue dEs close to the limiting
magnitude. For brighter and redder dEs, the total uncertainty decreases significantly, 
as deriving $\overline{M}_{\rm I}$ from $(V-I)$
is less uncertain for red objects and the measurement accuracy improves to about 0.1 mag for
the brightest simulated dEs. While the uncertainty in deriving $\overline{M}_{\rm I}$ from $(V-I)$
imposes a lower limit on the distance accuracy
when observing {\bf field} dEs, it allows rough age-metallicity estimates for blue {\bf cluster}
dEs, as most of the times the cluster is separated from background/foreground galaxies by 
significantly more than 1 mag in distance modulus.\\
We find that on average the measured SBF magnitude is 0.15 mag fainter than the simulated
one. A number of tests show that this is due to loss of fluctuation 
signal when recovering pixel-to-pixel fluctuations from a seeing convolved image. 
This average offset shows that much care must be taken when deriving absolute values 
like $H_{\rm 0}$ with the SBF-Method. When aiming at relative distances like for cluster 
membership determination, a bias is of no concern.\\  
By comparing real SBF data of Centaurus Cluster dEs with simulations tuned to reproduce the
real data, we find that our simulations do not overestimate the achievable S/N of the SBF method,
 but are consistent with real measurements. Therefore the statements about limiting magnitudes 
for the technique made in this paper are reasonable and we would not expect a very different 
behaviour in real observations.\\
An ideal application of the SBF technique would be a deep {\it and} wide field survey 
of several nearby clusters such as Fornax, Virgo or Doradus. With the arrival of 
wide field cameras on large telescopes (Suprime cam on the Subaru 
telescope or IMACS on Magellan), this is a very promising possibility to determine 
well the very faint end of the galaxy luminosity function in nearby clusters.\\ 
 
\label{conclusions}
\acknowledgements
We thank the referee N. Trentham for his very helpful comments.
SM was supported by DAAD PhD grant Kennziffer D/01/35298. LI
acknowledges support by FONDAP Centro de Astrof\'\i sica No. 15010003. This work is 
partially based on observations obtained at the European Southern Observatory,
Chile (Observing Programme 67.A--0358).\\

\enddocument